%% file: neurips_2026.tex
\documentclass{article}

\usepackage[preprint]{neurips_2026}

\usepackage[utf8]{inputenc} 
\usepackage[T1]{fontenc}    
\usepackage{hyperref}       
\usepackage{url}            
\usepackage{booktabs}       
\usepackage{amsmath}        
\usepackage{amsfonts}       
\usepackage{nicefrac}       
\usepackage{microtype}      
\usepackage{xcolor}         

\usepackage{graphicx}
\usepackage{array}
\usepackage{multirow}
\newcolumntype{C}[1]{>{\centering\arraybackslash}p{#1}}
\newcolumntype{L}[1]{>{\arraybackslash}p{#1}}
\usepackage{booktabs}

\usepackage{xspace}
\newcommand{\criteriongen}{$\mathcal{G}$\xspace}

\usepackage{listings}

\lstdefinestyle{promptstyle}{
    basicstyle=\ttfamily\scriptsize,
    breaklines=true,
    breakatwhitespace=false,
    columns=fullflexible,
    frame=single,
    backgroundcolor=\color{gray!5},
    xleftmargin=1mm,
    xrightmargin=1mm,
    keepspaces=true,
    showstringspaces=false
}

\input{metadata}

\author{
Zhenyu Li$^1$, Aleksandar Cvejić$^1$, Zehui Chen$^2$, Peter Wonka$^1$ \\
$^1$KAUST, $^2$USTC \\
\small\url{https://zhyever.github.io/CriterAlign/} \\
}

\begin{document}

\maketitle

\input{sec/0_abstract}
\input{sec/1_intro}

\input{sec/2_related_work}
\input{sec/3_method}
\input{sec/4_experiments}
\input{sec/5_conclusion}

\newpage

\bibliographystyle{plainnat}
\bibliography{main}


\input{sec/x_appendix}


\newpage
\input{checklist.tex}

\end{document}

%% file: metadata.tex
\title{CriterAlign: Criterion-Centric Rationale Alignment for Code Preference Judging}

%% file: sec/0_abstract.tex
\begin{abstract}
Pairwise human preference prediction is central to evaluating code-generation systems, where quality often depends on task-specific trade-offs beyond functional correctness.
While rubric-based LLM judges improve interpretability by decomposing evaluation into explicit criteria, most existing pipelines remain pointwise: they score each response independently and derive preferences by comparing aggregated scores.
We show that this design is poorly matched to pairwise code preference prediction and can underperform a strong monolithic judge.
We propose \textsc{CriterAlign}, a criterion-centric framework that adapts rubric-based judging to pairwise preference evaluation through direct criterion-level pairwise judgments, tie-driven criterion refinement, swap-consistency filtering, and final pairwise synthesis.
We further introduce Human-Preference-Aligned Guidance (HPAG), synthesized offline from training examples by extracting recurring rationale gaps between human preferences and monolithic judge predictions, and injected into the criterion generator, criterion judge, and final judge.
On BigCodeReward, \textsc{CriterAlign} improves a Qwen2.5-VL-32B monolithic judge from $60.4\%$ to $66.3\%$ accuracy, with ablations confirming the contributions of pairwise criterion design and HPAG.
\end{abstract}

%% file: sec/1_intro.tex
\section{Introduction}
\label{sec:intro}

Large language models are increasingly used to generate code, user interfaces, visual programs, and interactive applications~\citep{humaneval2021,swebench2024,webdevjudge2025,bigcodearena2025}. 
As these systems become more capable, evaluating their outputs becomes increasingly difficult~\citep{llmservey2024l}. 
For many coding tasks, the quality often depends on task-specific trade-offs beyond functional correctness that is necessary but not sufficient~\citep{userstudy_codegen_2024}.
Thus, code evaluation is often a preference problem rather than a single correctness check: given the same instruction and two candidate solutions, the goal is to determine which response better satisfies human expectations~\citep{codejudgebench2025}.

Since human evaluation is expensive, recent work has turned to LLMs as automatic judges~\citep{judgelm2025,prometheus2024,llmasjudgesurvey2024,llmasjudgesurvey22024}. 
However, monolithic LLM judges remain brittle for code preference prediction. 
A single-step judge must implicitly decide which task-specific criteria matter, compare the two candidate responses under those criteria, and aggregate the evidence into an overall preference~\citep{mctsjudge2025}. 
Because this reasoning process is hidden, the judge can over-emphasize superficial cues such as verbosity or formatting, under-use execution or visual evidence, or apply a generic notion of quality that does not match the specific coding task~\citep{rrd2026,rubricisall2025}.
Recent code-judge benchmarks further show that LLM judges can be sensitive to response ordering and evaluation format, raising concerns about their robustness in coding scenarios~\citep{codejudgebench2025}.

A natural alternative is criterion-based, or rubric-based, judging.
Instead of directly asking a model for a holistic preference, rubric-based methods generate explicit evaluation criteria and judge responses under those criteria. 
This decomposition improves interpretability and encourages the judge to consider multiple dimensions of quality. 
Recent methods such as LLM Rubrics, Chasing the Tail, and Recursive Rubric Decomposition (RRD) improve rubric generation by grounding criteria in responses, recursively decomposing broad rubrics, filtering redundant criteria, or reweighting correlated rubric signals~\citep{chasingtail2025,rrd2026}. 
However, most of these pipelines remain \emph{pointwise}: each response is evaluated independently under each criterion, and the final preference is derived by comparing scores. 
This is poorly matched to pairwise human preference prediction, where the key evidence is often comparative~\citep{codejudgebench2025}.

\begin{figure}
    \centering
    \includegraphics[width=0.95\linewidth]{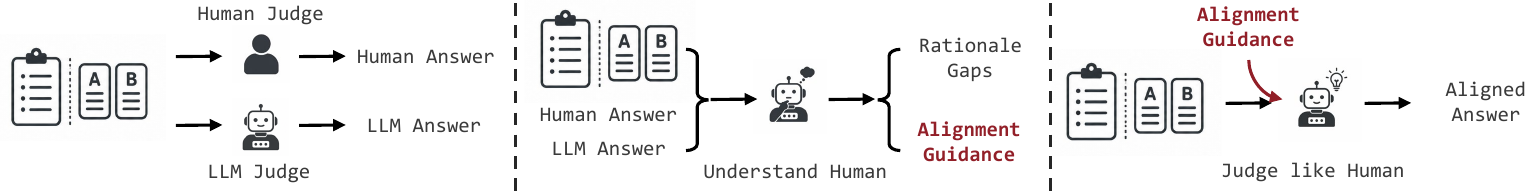}
    \caption{
\textbf{Learning human preference guidance for code judging.}
Given the same coding task and two candidate responses, human judges and LLM judges may produce different preference decisions.
\textsc{CriterAlign} analyzes the human and LLM decisions on training cases, summarizes rationale gaps into an alignment guidance, and injects the guidance into the judge at inference time.
This enables the LLM judge to better match human preferences while keeping the judge model frozen.
}
    \label{fig:teaser}
\end{figure}

In this paper, we ask: \emph{How can criterion-based judging be adapted to pairwise human preference prediction for code generation?} 
We propose \textsc{CriterAlign}, a criterion-centric judging framework that aligns criterion generation, criterion judge, and final judge with human pairwise preferences. 
Given a coding instruction, two candidate responses, and available execution artifacts, \textsc{CriterAlign} first generates task-specific criteria. 
Instead of assigning independent pointwise scores to each response, it asks a criterion judge to produce a pairwise verdict for each criterion: $v \in \{A, B, \text{tie}, \text{insufficient}\}$. 
Tied criteria are further refined through batched tie-driven criterion refinement, which replaces coarse non-discriminative criteria with finer comparative sub-criteria.
Motivated by the response-order sensitivity observed in pairwise code judging~\citep{codejudgebench2025}, we introduce swap-consistency criterion filtering, which removes criterion-level evidence whose verdict is not stable under swapping the candidate order.

Beyond adapting the judging pipeline to pairwise comparison, we further align it with human preferences through Human-Preference-Aligned Guidance (HPAG). 
Prior work has shown that natural-language principles and rubrics can serve as effective alignment or evaluation signals, guiding model behavior and fine-grained judging~\citep{cai2022,prometheus2024,codeultrafeedback2025}. 
HPAG follows this direction but derives the guidance from observed gaps between human preferences and judge predictions. 
Specifically, as shown in Figure~\ref{fig:teaser}, HPAG is synthesized offline from the training split only. 
A guidance synthesizer compares human decisions with monolithic judge predictions and extracts recurring rationale gaps: dimensions that humans care about but the judge tends to miss, over-emphasize, or under-emphasize. 
The resulting guidance is distilled into compact general and category-level preference principles and injected into the criterion generator, criterion judge, and final judge.
This enables reasoning with explicit human-aligned preference guidance at inference time.

We evaluate \textsc{CriterAlign} on BigCodeReward~\citep{bigcodearena2025}, a benchmark for pairwise code preference prediction. 
A monolithic Qwen2.5-VL-32B judge obtains $60.4\%$ accuracy on the held-out validation split. 
Directly adapting existing pointwise criterion-generation baselines performs worse than this monolithic judge, indicating that more criteria alone do not solve the problem. 
Consistent with the pairwise-over-pointwise finding in CodeJudgeBench~\citep{codejudgebench2025}, converting criterion judging from pointwise scoring to pairwise comparison already improves performance.

Our contributions are summarized as follows:
\begin{itemize}
    \item We identify and empirically validate a mismatch between pointwise criterion-generation pipelines and pairwise code preference prediction: independently scoring each response under generated criteria can underperform a strong monolithic pairwise judge.

    \item We propose \textsc{CriterAlign}, a pairwise criterion-centric judging framework that combines criterion-level verdicts, batched tie-driven criterion refinement, swap-consistency criterion filtering, and HPAG distilled from human--judge preference and rationale gaps.

    \item We demonstrate consistent gains on BigCodeReward, where \textsc{CriterAlign} improves a Qwen2.5-VL-32B~\cite{qwen25} judge from $60.4\%$ to $66.3\%$, with ablations isolating the contributions of pairwise adaptation, criterion filtering, and HPAG.
\end{itemize}

%% file: sec/2_related_work.tex
\section{Related Work}
\label{sec:related_work}

\paragraph{LLM-as-a-judge for code evaluation.}
LLMs are increasingly used not only to generate code but also to evaluate generated code~\citep{llmasjudgesurvey2024,llmasjudgesurvey22024,mctsjudge2025,codevisionary2025,rubricisall2025,codefavor2024,vibechecker2025,copilotarena2025,codeultrafeedback2025}. 
Early work studies LLM-based code judging for correctness, code understanding, repair, or educational grading, often in pointwise settings where a single response is scored or classified~\citep{codejudge2024,codejudgeeval2024}. 
Recent benchmarks move closer to preference-based evaluation. 
BigCodeArena and BigCodeReward collect pairwise human preferences for diverse coding tasks with execution or visual artifacts~\citep{bigcodearena2025}; WebDevJudge focuses on visually grounded web-development preferences~\citep{webdevjudge2025}; CodeJudgeBench studies LLM-as-a-judge across code generation, repair, and unit-test generation in a monolithic non-rubric setting and finds that pairwise prompting outperforms scalar pointwise judging while remaining sensitive to response order~\citep{codejudgebench2025}.
These benchmarks show that code evaluation is not only about functional correctness, but also about comparing task-specific qualities that affect human preference. 
Our work builds on this pairwise preference and studies how to make criterion-based judges aligned with such human comparisons.

\paragraph{Rubric- and criterion-based judges.}
A common way to improve LLM judges is to make evaluation criteria explicit. 
Manual or calibrated rubric prompting methods use natural-language rubrics to guide fine-grained assessment~\citep{geval2023,prometheus2024,prometheus2_2024,hdeval2024}. 
More recent work automates rubric construction through checklist generation, criterion decomposition, or adaptive rubric refinement~\citep{tickeval2024,autorubric2026,rulers2026,iruler2026,learningtojudge2026}. 
Another line uses rubrics as reward signals or post-training supervision~\citep{rar2025,onlinerubrics2025,openrubrics2025,rrd2026,openrs2026,cdrrm2026}. 
These methods show that generated criteria can be improved through recursive refinement, response-conditioned instantiation, preference-contrast synthesis, or correlation-aware weighting. 
However, they are mainly developed for general open-ended judging or reward modeling, and most do not directly target code-specific pairwise human preference prediction. 
\textsc{CriterAlign} instead focuses on code-specific pairwise human preference prediction, using criterion-based judging as a scaffold for human-aligned comparison rather than pointwise scoring.

\paragraph{Biases and diagnostics in LLM judging.}
LLM judges are known to exhibit biases from response order, verbosity, style, formatting, self-preference, and other superficial cues~\citep{selfpreferencebias2024,dontjudgecode2025,codejudgebench2025}. 
These issues are especially important for code preference prediction, where two responses may differ in both executable behavior and presentation. 
Prior diagnostic work studies when LLM judges disagree with humans or when judge predictions should be escalated to human review~\citep{trace2026,trustorescalate2024}. 
TRACE is particularly relevant because it compares developer and LLM biases in realistic code evaluation and shows that humans and LLM judges can systematically weight evaluation dimensions differently~\citep{trace2026}. 
Our work uses these observations as intervention targets: swap-consistency criterion filtering removes criterion-level evidence that is unstable under response-order swaps, while HPAG distills human--judge divergences into explicit preference guidance.

\paragraph{Guidance from human preferences.}
Several lines of work show that natural-language principles, rubrics, critiques, or feedback can guide model behavior and evaluation~\citep{cai2022,prometheus2024,codeultrafeedback2025}. 
Constitutional-style methods encode desired behavior as natural-language principles, while critique- or rubric-conditioned evaluators use textual guidance to structure assessment~\citep{cai2022,cloud2024,genrm2024,salmon2023}. 
Closer to our setting, preference-derived guidance can be distilled from annotations and reused as inference-time context for a frozen judge~\citep{icai2025}, while training-time approaches bake similar signals into model weights~\citep{spct2025}. 
\textsc{CriterAlign} follows the inference-time direction: HPAG is synthesized from training examples by comparing human preferences with monolithic judge predictions, then injected into the criterion generator, criterion judge, and final judge. 
Unlike general-purpose constitutions or training-time judge tuning, HPAG is designed to capture code-specific human--judge rationale gaps and to guide a pairwise criterion pipeline.

%% file: sec/3_method.tex
\section{\textsc{CriterAlign}}
\label{sec:method}

\begin{figure}
\definecolor{llmblue}{RGB}{27, 91, 150}
\definecolor{contriborange}{RGB}{198, 77, 17}
\definecolor{refinepurpletext}{RGB}{108, 80, 110} 

    \centering
    \includegraphics[width=0.99\linewidth]{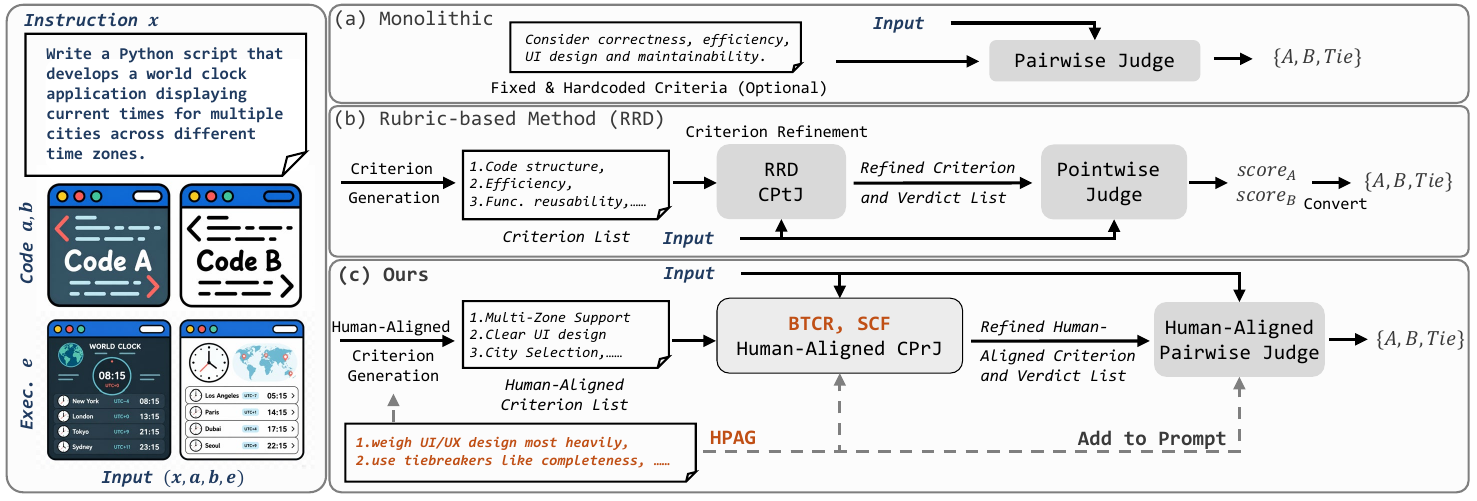}
    \caption{
\textbf{Inference Pipeline Comparison.}
Monolithic judges use fixed or implicit criteria, while rubric-based methods such as RRD~\citep{rrd2026} generate criteria but rely on pointwise criterion refinement and scoring.
\textsc{CriterAlign} synthesizes human-preference-aligned guidance (HPAG) offline from the training split and injects it into the pairwise rubric-based pipeline for human-aligned inference. \textcolor{contriborange}{\textbf{Orange}} highlights our main contributions. CPtJ and CPrJ represent criterion pointwise and pairwise judge, respectively. RRD denotes recursive rubric decomposition. BTCR and SCF denote our proposed batched tie-driven criterion refinement and swap-consistency filtering, respectively. Human-Preference-Aligned Guidance (HPAG) is synthesized offline from the training split in a separate guidance-synthesis stage; see Sec.~\ref{sec:method:hpag} for details.
}
    \label{fig:method}
\end{figure}

Figure~\ref{fig:method} illustrates the overall pipeline of \textsc{CriterAlign}. 
In this section, we first formulate pairwise human preference prediction for code generation in Sec.~\ref{sec:method:setup}. 
We then review Recursive Rubric Decomposition (RRD) in Sec.~\ref{sec:method:rrd}, which motivates criterion-based judging but is originally built around pointwise rubric scoring. 
Finally, we present \textsc{CriterAlign} in Sec.~\ref{sec:method:pipeline}. 
The pipeline first adapts criterion decomposition to pairwise preference prediction through pairwise criterion judging, batched tie-driven refinement, and swap-consistency filtering. 
It then incorporates human-preference-aligned guidance to align criterion generation, criterion judging, and final synthesis with human preference rationales.

\subsection{Problem Setup}
\label{sec:method:setup}

We study pairwise human preference prediction for code generation. 
Each instance is denoted as $(x, a, b, e, y)$, where $x$ is the coding instruction, $a$ and $b$ are two candidate code responses, $e$ denotes optional execution evidence such as program outputs, screenshots, or error traces, and $y$ is the human preference label. 
Following the BigCodeReward protocol, we map the original label space to $\mathcal{Y}=\{A,B,\text{tie}\}$ by merging ambiguous failure cases into \texttt{tie}. 
A judge $J$ takes $(x,a,b,e)$ as input and predicts $\hat{y}\in\mathcal{Y}$. 
The goal is to maximize agreement between $\hat{y}$ and the human preference $y$ on held-out examples.

A monolithic judge directly predicts $\hat{y}$ in a single step. 
In contrast, \textsc{CriterAlign} decomposes the decision into task-specific criteria. 
Given an instance $(x,a,b,e)$, it first generates a set of criteria $\mathcal{C}=\{c_i\}_{i=1}^{m}$, then obtains a criterion-level pairwise verdict for each criterion, and finally synthesizes these criterion-level judgments into an overall preference prediction. 
This design is motivated by the observation that human preferences over code are multi-dimensional and task-dependent: depending on the instruction, humans may emphasize visual fidelity, robustness, simplicity, maintainability, or other latent requirements.

\subsection{Preliminary: Recursive Rubric Decomposition}
\label{sec:method:rrd}

Our method builds on the general idea of rubric-based judging, particularly Recursive Rubric Decomposition (RRD)~\citep{rrd2026}. 
In this setting, rubrics are not ground-truth annotations; they are generated by an LLM conditioned on the evaluation instance and used as intermediate evaluation aids. 
RRD starts from LLM-generated rubrics, recursively decomposes coarse rubrics into finer-grained criteria, filters misaligned or redundant rubrics, and then aggregates rubric-level signals with a weighting scheme. 
This framework is designed for open-ended evaluation, where explicit criteria can improve coverage and interpretability compared with a single holistic judgment.

However, directly applying RRD-style pipelines to our setting is suboptimal for two reasons. 
First, most rubric pipelines evaluate each candidate response independently under each criterion, producing pointwise satisfaction scores. 
Pairwise human preference prediction is different: the target is not whether each response individually satisfies a criterion, but which response better satisfies the criterion relative to the other. 
Second, pairwise judging introduces its own failure modes, especially position sensitivity: a judge may prefer the first-presented response even when the candidate order is swapped. 
\textsc{CriterAlign} therefore adapts criterion decomposition to the pairwise setting and adds a criterion-level consistency filter specifically designed for pairwise comparison.

\subsection{\textsc{CriterAlign} Pipeline}
\label{sec:method:pipeline}

\paragraph{Pairwise Criterion Judging.}
\label{sec:method:pairwise}

For each instance, the criterion generator produces a task-specific criterion set $\mathcal{C}=\{c_i\}_{i=1}^{m}$ conditioned on the instruction, the two candidate responses, and available execution evidence. 
Each criterion is required to be atomic, comparative, and judgeable from the provided evidence. 
Rather than asking whether response $a$ and response $b$ separately satisfy $c_i$, we ask the criterion judge to compare them directly under $c_i$:
\[
v_i = J_{\mathrm{crit}}(x,a,b,e,c_i),
\]
where
\[
v_i \in \{A, B, \text{tie}, \text{insufficient}\}.
\]
Here, $A$ means that response $a$ better satisfies criterion $c_i$, $B$ means that response $b$ better satisfies it, \texttt{tie} means that the two responses are comparable under this criterion, and \texttt{insufficient} means that the available evidence is not enough to judge the criterion reliably.

This pairwise reformulation changes both criterion evaluation and final aggregation. 
In pointwise rubric pipelines such as RRD, each response is independently scored under each rubric, and the final preference is obtained by comparing weighted sums of rubric scores. 
In \textsc{CriterAlign}, each criterion instead produces direct comparative evidence about the two candidate responses, and the final judge synthesizes these criterion-level verdicts into an overall $A/B/\text{tie}$ decision. 
We use an LLM judge rather than a fixed weighted-sum rule because human preferences often involve non-uniform trade-offs across criteria: a large correctness gap may dominate several minor style advantages, while visual fidelity may be decisive in web-design tasks~\cite{codejudgebench2025}.

\paragraph{Batched Tie-Driven Criterion Refinement (BTCR).}
\label{sec:method:batched-refinement}

After initial pairwise criterion judging, some criteria receive a \texttt{tie} verdict. 
Rather than treating every tie as final evidence, we refine tied criteria because a tie may indicate that the criterion is too coarse to expose the concrete difference between the two responses. 

Given the current criteria and their pairwise verdicts, the refiner identifies criteria judged as \texttt{tie}. 
For each tied parent criterion, it generates up to two finer-grained sub-criteria conditioned on the instruction, candidate responses, available execution evidence, the parent criterion, its tie rationale, and the other existing criteria. 
Following RRD, we filter generated sub-criteria that are redundant with or conflicting with existing criteria. 
Accepted sub-criteria are then re-judged by the same pairwise criterion judge and treated as independent criterion-level evidence. 
If different children favor different responses, we keep both verdicts as fine-grained trade-off evidence and let the final judge synthesize them into the overall preference.

This refinement is triggered by pairwise ties, unlike RRD-style decomposition based on pointwise satisfaction over multiple sampled responses. 
Thus, its goal is not to indefinitely expand the criterion set, but to replace non-discriminative tied evidence with more fine-grained comparative evidence. 
For efficiency, we batch decomposition and filtering: all tied criteria in one iteration are decomposed in a single structured call, and candidate sub-criteria are filtered with one batch redundancy check and one batch conflict check.

\paragraph{Swap-Consistency Criterion Filtering (SCF).}
\label{sec:method:scf}

Pairwise judges are known to be sensitive to candidate order. 
In our setting, this bias can occur not only in the final preference decision but also at the criterion level. 
A criterion verdict may favor response $a$ when the input order is $(a,b)$, but fail to favor the same response when the order is swapped to $(b,a)$. 
Such position-sensitive criterion judgments are unreliable evidence for final aggregation.

We introduce \emph{Swap-Consistency Criterion Filtering} (SCF), a pairwise-specific filtering step applied after tie-driven criterion refinement. 
Let $\mathcal{C}'$ denote the current criterion set after refinement, and let
\[
v_i^{\rightarrow} = J_{\mathrm{crit}}(x,a,b,e,c_i)
\]
be the already obtained criterion verdict for criterion $c_i \in \mathcal{C}'$. 
SCF re-judges the same criterion under the swapped candidate order:
\[
v_i^{\leftarrow} = J_{\mathrm{crit}}(x,b,a,e^{\leftarrow},c_i),
\]
where $e^{\leftarrow}$ denotes the execution evidence with the artifacts of responses $a$ and $b$ swapped accordingly.

We then map the swapped verdict back to the original candidate order through a swap operator $\pi(\cdot)$:
\[
\pi(A)=B,\quad \pi(B)=A,\quad 
\pi(\text{tie})=\text{tie},\quad 
\pi(\text{insuf.})=\text{insuf.}.
\]
A criterion-level judgment is kept only when
\[
v_i^{\rightarrow} = \pi(v_i^{\leftarrow}).
\]
Otherwise, the corresponding criterion--judgment pair is discarded before final synthesis.

SCF differs from holistic order-swap debiasing. 
Instead of discarding the entire pairwise example when the final answer is inconsistent, SCF operates at the criterion level. 
This allows the pipeline to remove unstable evidence while retaining other criteria whose pairwise verdicts are order-consistent. 
As a result, SCF serves both as a debiasing mechanism and as a quality filter for criterion-level evidence.

\paragraph{Human-Preference-Aligned Guidance (HPAG).}
\label{sec:method:hpag}

Criterion decomposition alone does not guarantee human alignment. 
Even when criteria are valid and pairwise-consistent, the judge may still emphasize dimensions differently from human annotators.  
We address this mismatch with \emph{Human-Preference-Aligned Guidance}.

HPAG is synthesized \textit{\textbf{offline}} using only the \textit{\textbf{training split}} (20\% of the entire dataset). 
For each training example, we construct a guidance-synthesis record in three steps. 
First, we run a monolithic LLM judge on the input $(x,a,b,e)$ and ask it to output both a pairwise preference prediction and a free-form rationale. 
Second, to make the human side explicit, we ask a rationale-analysis LLM to produce a concise hypothesis for the human-vote rationale based only on the training example and its human preference labels. 
Third, we package these signals into a structured record:
\[
r = (x, a, b, \hat{y}_{\mathrm{mono}}, s_{\mathrm{mono}}, y, y_{\mathrm{aspect}}, s_{\mathrm{human}}),
\]
where $\hat{y}_{\mathrm{mono}}$ and $s_{\mathrm{mono}}$ denote the monolithic judge prediction and rationale, $y$ is the human vote, $y_{\mathrm{aspect}}$ denotes aspect-level labels, and $s_{\mathrm{human}}$ is the inferred human-vote rationale.

The guidance synthesizer takes the concatenated list of these training records as input and summarizes recurring gaps between the monolithic judge and human preferences. 
These gaps describe dimensions that humans consistently care about but the judge tends to miss, over-emphasize, or under-emphasize. 
The output is a compact natural-language guidance artifact that is frozen after offline construction and never updated using validation examples. 
We present the synthesized HPAG in Appendix~\ref{app:hpag}.

We synthesize two types of guidance. 
\emph{General guidance} summarizes dataset-level preference principles shared across tasks. 
\emph{Category-level guidance} captures preference patterns for different task families, such as web design, game development, diagram creation, creative coding, scientific computing, and problem solving. 
At inference time, HPAG is injected into the criterion generator, criterion judge, and final judge. 
For the criterion generator, it encourages criteria that reflect human-relevant preference dimensions. 
For the criterion judge, it reminds the judge how humans tend to compare responses under each criterion. 
For the final judge, it guides the synthesis of criterion-level evidence toward human preference.

\paragraph{Final Judge.}
\label{sec:method:final}

After batched tie-driven refinement and swap-consistency criterion filtering, the remaining criterion--verdict pairs form a set of reliable pairwise evidence:
\[
\mathcal{E}=\{(c_i,v_i): v_i^{\rightarrow}=\pi(v_i^{\leftarrow})\}.
\]
The final judge receives the original instance, retained criterion-level evidence, and final-stage HPAG:
\[
\hat{y}=J_{\mathrm{final}}(x,a,b,e,\mathcal{E};g_{\mathrm{final}}),
\quad
\hat{y}\in\{A,B,\text{tie}\}.
\]

Overall, \textsc{CriterAlign} adapts criterion-centric judging to pairwise code preference prediction through four key changes: pairwise criterion verdicts replace pointwise satisfaction scores, BTCR replaces coarse tied criteria with finer comparative evidence, SCF removes order-sensitive criterion evidence, and HPAG aligns the criterion generator, criterion judge, and final judge with human preference rationales.

%% file: sec/4_experiments.tex
\section{Experiments}
\label{sec:experiments}

\subsection{Setup}
\label{sec:experiments:setup}

\paragraph{Dataset.}
We mainly evaluate our method on BigCodeReward, a benchmark for pairwise human preference prediction in code generation~\citep{bigcodearena2025}. 
Each example contains a coding instruction, two candidate solutions, available execution or visual artifacts, and human preference labels. 
The instructions cover diverse code-generation scenarios, including web design, game development, diagram creation, creative coding, scientific computing, and general problem solving. To avoid information leakage, we randomly split BigCodeReward into a training split and a held-out validation split. 
We use the training split only to synthesize the human-preference-aligned guidance, and report all main results on the held-out validation split. 
Following our protocol, the validation split contains $3{,}785$ examples, corresponding to approximately $80\%$ of the data, while the remaining examples are used for guidance synthesis. 
Unless otherwise specified, we report overall preference accuracy on the validation split.


\paragraph{Implementation details}
Our default judge is Qwen2.5-VL-32B~\citep{qwen25}. 
The monolithic baseline directly asks the judge to compare the two candidate responses and predict the human-preferred answer. 
\textsc{CriterAlign} uses the same judge backbone but decomposes the decision process into multiple stages. 
Unless otherwise stated, the default human-preference-aligned guidance synthesizer is Claude Sonnet 4.6~\citep{sonnet2024}, which supports a long context window that can consume the full training split for guidance synthesis. 
For synthesizers with shorter context windows, we downsample the training split to the largest subset that fits within the context budget, ensuring that all guidance is still synthesized only from training examples. 
We serve all open-weight models with vLLM~\citep{vllm2023} on a node with $4\times$ H200 GPUs, using the same decoding settings across methods unless otherwise specified; proprietary synthesizers are accessed through their API.
We provide all prompt templates and synthesized guidance in Appendix~\ref{app:prompts} and Appendix~\ref{app:hpag}, respectively.

\input{tables/main_table}

\subsection{Main results}
\label{sec:experiments:main}

Our main comparisons are against criterion-generation-based pipelines, including LLM Rubrics~\citep{rrd2026}, Rubric Is All You Need~\citep{rubricisall2025}, Chasing the Tail~\citep{chasingtail2025}, and RRD~\citep{rrd2026}.
We reproduced and report their results on the BigCodeArena.
Table~\ref{tab:main} summarizes the main results. 
On BigCodeReward validation, the monolithic Qwen2.5-VL-32B judge obtains $60.4\%$ accuracy. 
Existing criterion-generation-based baselines~\citep{rrd2026,rubricisall2025,chasingtail2025,rrd2026} underperform this monolithic judge by a large margin. 
This result highlights an important mismatch: although criterion decomposition is attractive for open-ended evaluation, directly applying existing pointwise criterion pipelines to pairwise code preference prediction can hurt rather than help. 
In contrast, \textsc{CriterAlign} reaches $66.3\%$ accuracy, improving over the monolithic baseline by $+5.9$ points and over the strongest reproduced criterion-generation baseline~\citep{rrd2026} by $+11.3$ points. 
This improvement indicates the effectiveness of our design choices. We show one detailed case in Appendix~\ref{app:case_rule30}.

\input{tables/ablation_overall}

\subsection{Component ablation}
\label{sec:experiments:ablation}

Table~\ref{tab:ablation} decomposes the improvement of \textsc{CriterAlign}. 
Starting from the monolithic Qwen2.5-VL-32B baseline at $60.4\%$, pointwise RRD drops to $55.0\%$, confirming that a direct pointwise rubric pipeline is not a strong baseline for this task. 
The largest recovery comes from adapting criterion judging to the pairwise setting. 
Replacing pointwise satisfaction scoring with pairwise criterion comparison raises accuracy from $55.0\%$ to $61.8\%$, a $+6.8$ point gain over pointwise RRD and a $+1.4$ point improvement over the monolithic judge. 
We provide a more detailed pointwise-to-pairwise adaptation analysis in Appendix~\ref{sec:experiments:point2pair}. 
This result suggests that criteria become useful when they provide comparative evidence about which response better satisfies each criterion.
The remaining components further improve this pairwise criterion pipeline. 
Batched tie-driven criterion refinement increases accuracy from $61.8\%$ to $62.2\%$ by replacing coarse tied criteria with finer comparative evidence. 
Adding global HPAG improves accuracy to $63.6\%$, showing that preference-aligned guidance helps the criterion generator and criterion judge better reflect human-relevant dimensions. 
Swap-consistency filtering further raises accuracy to $64.6\%$, indicating that filtering order-sensitive criterion evidence improves robustness. 
Finally, category-level HPAG gives the full \textsc{CriterAlign} model at $66.3\%$, suggesting that category-specific preference guidance provides complementary gains beyond global guidance.
Overall, the ablation shows that \textsc{CriterAlign}'s improvement is cumulative rather than driven by a single prompt change: pairwise criterion judging fixes the main pointwise mismatch, BTCR refines non-discriminative ties, SCF removes unstable criterion evidence, and HPAG aligns the criterion generator, criterion judge, and final judge with human preferences.

\input{tables/ablation_more_details}

\subsection{Pipeline-integrity controls}
\label{sec:experiments:controls}

Table~\ref{tab:pipeline_controls} provides additional controls for HPAG and its injection points. 
Adding global HPAG to the monolithic judge improves accuracy from $60.4\%$ to $63.3\%$; further adding category-level HPAG raises it to $64.4\%$, showing that HPAG helps even without criterion decomposition.
However, the full \textsc{CriterAlign} pipeline reaches $66.3\%$, indicating that guidance alone does not explain the entire gain.
Category-guidance controls test whether gains come from category-level guidance itself, rather than category labels or prompt slots. 
Removing category-level HPAG gives $64.6\%$, while empty category guidance gives $64.9\%$, suggesting that the category scaffold alone provides only a small gain. 
Finally, HPAG is most effective when injected throughout the pipeline. 
Without HPAG, the decomposed pipeline obtains $62.2\%$; injecting HPAG only into criterion generation, only into criterion judging, and into both stages gives $63.4\%$, $64.6\%$, and $65.2\%$, respectively. 
Adding HPAG to final synthesis reaches $66.3\%$, showing that best performance comes from aligning all three stages.

\input{tables/ablation_baselines}

\subsection{Generalization across judges and guidance synthesizers}
\label{sec:experiments:cross-model}

Table~\ref{tab:cross-model} evaluates whether the proposed guidance is tied to a specific judge or synthesizer. 
We vary the judge model across Kimi-VL-A3B~\citep{kimi2025}, Gemma-4-31B~\citep{gemma42026}, and Qwen2.5-VL-32B, and vary the model used to synthesize the guidance. 
Across all tested judges, \textsc{CriterAlign} improves over the matched monolithic baseline.
The relative gain is largest for the weakest judge. 
Kimi-VL-A3B obtains only $46.5\%$ as a monolithic judge, but improves to $56.9\%$ with self-synthesized guidance and to $60.7\%$ with Sonnet-synthesized guidance. 
Gemma-4-31B improves from $57.7\%$ to $64.5\%$ with self-synthesized guidance and $63.7\%$ with Sonnet-synthesized guidance. 
For Qwen2.5-VL-32B, the monolithic baseline is already stronger at $60.4\%$, but \textsc{CriterAlign} still improves it to $65.2\%$ with Qwen-synthesized guidance and $66.3\%$ with Sonnet-synthesized guidance.
These results show that \textsc{CriterAlign} is not tied to Qwen2.5-VL-32B or to a single proprietary guidance synthesizer. 
The gains from self-synthesized guidance suggest that HPAG provides a transferable alignment signal rather than a brittle model-specific prompt.

%% file: tables/main_table.tex
\begin{table}[t]
    \centering
    \caption{
    \textbf{Main results on BigCodeReward validation.}
    We compare \textsc{CriterAlign} with the monolithic Qwen2.5-VL-32B judge and criterion-generation-based baselines.
    Numbers are accuracy in percent; \textit{imp.} denotes improvement over the monolithic baseline.
    \textbf{Bold} marks the best result.
    }
    \label{tab:main}
    \vspace{1mm}
    \scalebox{0.88}{
    \begin{tabular}{L{8.0cm}|C{1.2cm}C{1.2cm}}
        \toprule
        \textbf{Method} & \textbf{Acc.} & \textbf{\textit{imp.}} \\
        \midrule
        Monolithic~\citep{qwen25} & $60.4$ & - \\
        \midrule
        Rubric Is All You Need~\citep{rubricisall2025} & $48.1$ & $-12.3$ \\
        Chasing the Tail~\citep{chasingtail2025} & $53.3$ & $-7.1$ \\
        LLM Rubrics~\cite{rrd2026} & $50.3$ & $-10.1$ \\
        RRD$_{\text{UW}}$~\cite{rrd2026} & $51.6$ & $-8.8$ \\ 
        RRD$_{\text{LLM}}$~\cite{rrd2026} & $55.0$ & $-5.4$ \\
        \midrule
        \textsc{CriterAlign} \textbf{(Ours)} & $\mathbf{66.3}$ & $\mathbf{+5.9}$ \\
        \bottomrule
    \end{tabular}
    }
\end{table}

%% file: tables/ablation_overall.tex
\begin{table}[t]
    \caption{
    \textbf{Component ablation on BigCodeReward validation.}
    Starting from pointwise RRD, we progressively adapt criterion-based judging to the pairwise preference setting and add each component of \textsc{CriterAlign}. 
    \textit{imp.} reports the cumulative improvement over the monolithic Qwen2.5-VL-32B.
    }
    \label{tab:ablation}
    \centering
    \scalebox{0.85}{
    \begin{tabular}{L{9.5cm}|C{1.3cm}C{1.3cm}}
    \toprule
    \textbf{Method} & \textbf{Acc.} & \textbf{\textit{imp.}} \\
    \midrule
    Qwen2.5-VL-32B (Monolithic) & $60.4$ & --- \\
    \midrule
    Pointwise RRD~\citep{rrd2026} & $55.0$ & $-5.4$ \\
    Pairwise RRD & $61.8$ & $+1.4$ \\
    \quad $+$ Batched tie-driven criterion refinement (BTCR) & $62.2$ & $+1.8$ \\
    \quad $+$ Global human-preference-aligned guidance (G-HPAG) & $63.6$ & $+3.2$ \\
    \quad $+$ Swap-consistency criterion filtering (SCF) & $64.6$ & $+4.2$ \\
    \quad $+$ Category-level human-preference-aligned guidance (C-HPAG) & $\mathbf{66.3}$ & $\mathbf{+5.9}$ \\
    \bottomrule
    \end{tabular}
    }
\end{table}

%% file: tables/ablation_more_details.tex
\begin{table}[t]
    \centering
    \caption{
    \textbf{Pipeline-integrity controls on BigCodeReward validation.}
    Panel (a) tests whether HPAG helps outside the full criterion pipeline, whether category-level HPAG provides additional gains rather than merely exposing category labels.
    Panel (b) ablates where HPAG is injected in \textsc{CriterAlign}.
    }
    \label{tab:pipeline_controls}

    \vspace{-2.0mm}

    \begin{minipage}[t]{0.49\linewidth}
        \centering
        \textbf{\small(a) Guidance controls\vspace{0.5mm}}
        \scalebox{0.85}{
        \begin{tabular}{L{5.0cm}|C{1.5cm}}
            \toprule
            \textbf{Variant} & \textbf{Acc.} \\
            \midrule
            Monolithic baseline & 60.4 \\
            + Global HPAG & 63.3 \\
            + Category-level HPAG & 64.4 \\
            \midrule
            w/o Category-level HPAG & 64.6 \\
            Empty category guidance & 64.9 \\
            Full \textsc{CriterAlign} \textbf{(Ours)} & \textbf{66.3} \\
            \bottomrule
        \end{tabular}
        }
    \end{minipage}
    \hfill
    \begin{minipage}[t]{0.49\linewidth}
        \centering
        \textbf{\small(b) Per-stage HPAG ablation\vspace{0.5mm}}
        \scalebox{0.85}{
        \begin{tabular}{C{1.8cm}C{1.8cm}C{1.8cm}|C{1.0cm}}
            \toprule
            \textbf{Gen.} & \textbf{Crit. Judge} & \textbf{Final} & \textbf{Acc.} \\
            \midrule
             &  &  & 62.2 \\
            \checkmark &  &  & 63.4 \\
             & \checkmark &  & 64.6 \\
            \checkmark & \checkmark &  & 65.2 \\
            \midrule
            \checkmark & \checkmark & \checkmark & $\mathbf{66.3}$ \\
            \bottomrule
        \end{tabular}
        }
        \vspace{0.5mm}
        
        {\small\textbf{Ours}: HPAG injected into all three stages.}
    \end{minipage}
\end{table}

%% file: tables/ablation_baselines.tex
\begin{table}[t]
    \caption{
    \textbf{Generalization across judges and guidance synthesisers.}
    We vary both the judge model and the model used to synthesize HPAG.
    \textit{Mono.} and \textit{Ours} denote the accuracy of the matched monolithic baseline and \textsc{CriterAlign}, respectively.
    \textit{imp.} reports the improvement over the matched monolithic baseline.
    Numbers are accuracy in percent.
    } 
    \label{tab:cross-model}
    \centering
    \scalebox{0.8}{
    \begin{tabular}{L{3.2cm}L{3.2cm}|C{1.8cm}C{1.8cm}C{1.8cm}}
    \toprule
    \textbf{Judge} & \textbf{Guidance synthesiser} & \textbf{Mono.} & \textbf{Ours} & \textbf{\textit{imp.}} \\
    \midrule
    \multirow{2}{*}{Kimi-VL-A3B} 
        & Kimi-VL-A3B  & \multirow{2}{*}{$46.5$} & $56.9$ & $+10.4$ \\
        & Sonnet 4.6   & & $60.7$ & $+14.2$ \\
    \midrule
    \multirow{2}{*}{Gemma-4-31B}   
        & Gemma-4-31B & \multirow{2}{*}{$57.7$} & $64.5$ & $+6.8$ \\
        & Sonnet 4.6  & & $63.7$ & $+6.0$ \\
    \midrule
    \multirow{2}{*}{Qwen2.5-VL-32B}    
        & Qwen2.5-VL-32B & \multirow{2}{*}{$60.4$} & $65.2$ & $+4.8$ \\
        & Sonnet 4.6 & & $66.3$ & $+5.9$ \\
    \bottomrule
    \end{tabular}
    }
\end{table}

%% file: sec/5_conclusion.tex
\section{Conclusion}
\label{sec:conclusion}

We presented \textsc{CriterAlign}, a criterion-centric framework for pairwise code preference prediction. 
Instead of independently scoring each response under pointwise rubrics, \textsc{CriterAlign} adapts criterion decomposition to pairwise human preference prediction through criterion-level A/B/tie judgments, tie-driven refinement, swap-consistency filtering, and human-preference-aligned guidance. 
On BigCodeReward, \textsc{CriterAlign} improves a Qwen2.5-VL-32B monolithic judge from $60.4\%$ to $66.3\%$ accuracy and outperforms the strongest reproduced criterion-generation baseline by $+11.3$ points. 
Ablations show that the gains come from both pairwise criterion design and HPAG, while cross-model experiments demonstrate the generality of \textsc{CriterAlign} across different judge backbones and guidance synthesizers.
These results indicate that criterion-based judging for code preference prediction should not only generate more criteria, but also align how those criteria are compared and synthesized with human pairwise preferences.

%% file: sec/x_appendix.tex
\clearpage
\appendix

\section{More Experimental Results}

\subsection{From pointwise rubrics to pairwise preference judging}
\label{sec:experiments:point2pair}

Table~\ref{tab:point2pair} provides a detailed diagnostic comparison between pointwise and pairwise criterion-based judging. 
Most existing rubric pipelines are pointwise: they score each candidate response independently under each criterion and then compare aggregated scores. 
This formulation is natural for quality scoring, but it is poorly matched to pairwise human preference prediction, where the judge often needs to compare relative trade-offs directly. 
For example, one response may be more correct but less concise, while another may look visually better but fail to implement a required interaction.

The results support this diagnosis. 
LLM Rubrics with pointwise criterion judging obtains $47.5\%$ with uniform aggregation, $49.6\%$ with LLM-assigned weights, and $50.3\%$ when replacing the weighted sum with a final pairwise LLM judge. 
Pointwise RRD also remains below the monolithic baseline across prompt and aggregation variants, with its best variant reaching $55.0\%$. 
In contrast, adapting RRD to use a pairwise criterion judge raises accuracy to $61.8\%$, surpassing the monolithic judge by $+1.4$ points. 
This suggests that criteria are useful only when the criterion-level judgment itself matches the pairwise preference target.

\textsc{CriterAlign} builds on this pairwise adaptation by replacing RRD's recursive refinement with batched tie-driven refinement, applying swap-consistency criterion filtering, and injecting HPAG into both criterion judging and final synthesis. 
The full pipeline reaches $66.3\%$, indicating that pairwise conversion is necessary but not sufficient: the best result comes from combining pairwise criterion judgments with robust criterion filtering and human-preference-aligned guidance.

\input{tables/point2pair}

\subsection{Effectiveness of Swap-Consistency Criterion Filtering}
\label{app:scf-analysis}

Swap-Consistency Criterion Filtering (SCF) is designed to remove criterion-level judgments that are sensitive to the order of the two candidate responses.
To verify whether SCF indeed reduces position bias, we analyze the criterion judgments before and after SCF on the  validation set.

Table~\ref{tab:scf_bias_analysis} reports the A/B/tie distribution at two levels.
At the criterion level, we count all criterion-level verdicts across validation examples.
At the sample level, we compute which side wins more criteria for each example.
Before SCF, the criterion judge exhibits a strong A-position bias: A receives $44.4\%$ of criterion verdicts while B receives only $31.5\%$, a $+12.9$ percentage-point skew.
After SCF, the distribution becomes nearly balanced, with $40.3\%$ A and $42.0\%$ B.
A similar pattern appears at the sample level: the pre-SCF majority distribution favors A by $+17.3$ points, while the post-SCF distribution has only a $-0.9$ point skew, close to the human reference skew of $+1.6$ points.
This indicates that a large fraction of the apparent criterion evidence before filtering is position-sensitive rather than semantically reliable.

\begin{table}[t]
    \centering
    \caption{
    \textbf{Effect of Swap-Consistency Criterion Filtering (SCF) on position bias.}
    We compare the A/B/tie distribution before and after SCF on the validation set.
    Criterion-level statistics count all criterion judgments, while sample-level statistics count which side wins more criteria per example.
    SCF substantially reduces the pre-filtering A-position bias and brings the sample-level majority distribution close to the human reference.
    }
    \label{tab:scf_bias_analysis}
    \scalebox{0.9}{
    \begin{tabular}{L{4.2cm}|C{1.3cm}C{1.3cm}C{1.3cm}|C{1.5cm}C{1.3cm}}
        \toprule
        \textbf{Stage} & \textbf{A} & \textbf{B} & \textbf{Tie} & \textbf{A-B (pp)} & \textbf{A/B} \\
        \midrule
        \multicolumn{6}{l}{\emph{Criterion-level judgment distribution}} \\
        Pre-SCF $(n=75{,}177)$  & $44.4$ & $31.5$ & $24.1$ & $+12.9$ & $1.41$ \\
        Post-SCF $(n=55{,}511)$ & $40.3$ & $42.0$ & $17.7$ & $-1.7$ & $0.96$ \\
        \midrule
        \multicolumn{6}{l}{\emph{Sample-level criterion-majority distribution}} \\
        Pre-SCF                 & $56.4$ & $39.1$ & $4.5$  & $+17.3$ & $1.44$ \\
        Post-SCF                & $46.1$ & $47.0$ & $6.9$  & $-0.9$  & $0.98$ \\
        Human reference          & $45.0$ & $43.4$ & $11.6$ & $+1.6$  & $1.04$ \\
        \bottomrule
    \end{tabular}
    }
\end{table}

SCF is also active on most validation examples.
In this run, $92.9\%$ of validation samples have at least one criterion removed by SCF.
Overall, SCF removes $25{,}464$ position-sensitive criterion judgments, corresponding to approximately $34\%$ of all initial criterion judgments, and reduces the criterion pool from $75{,}177$ to $55{,}511$.
These results support the design choice of filtering at the criterion level: rather than discarding an entire pairwise example, SCF selectively removes unstable evidence while preserving order-consistent criterion judgments.

\subsection{Efficiency of Batched Tie-Driven Criterion Refinement}
\label{app:btcr-efficiency}

Batched Tie-Driven Criterion Refinement (BTCR) refines criteria that receive a \texttt{tie} verdict after the initial pairwise criterion judging step.
A straightforward implementation, following the original RRD-style refinement loop~\citep{rrd2026}, would process each tied parent criterion separately: decompose each tied criterion, check each generated sub-criterion for redundancy, check each non-redundant sub-criterion for conflict, and then re-judge accepted sub-criteria.
This per-criterion loop can require many LLM calls because each tied criterion typically produces multiple candidate sub-criteria.

Our implementation batches the expensive refinement operations within each refinement iteration.
All tied criteria in the current iteration are decomposed in one structured LLM call.
All generated sub-criteria are then checked with one batch redundancy call and one batch conflict call.
Thus, if an iteration contains $N_{\mathrm{tie}}$ tied parent criteria, $N_{\mathrm{sub}}$ generated sub-criteria, and $N_{\mathrm{nonred}}$ non-redundant sub-criteria, the plain loop requires
\[
N_{\mathrm{tie}} + N_{\mathrm{sub}} + N_{\mathrm{nonred}}
\]
LLM calls for decomposition and filtering, while BTCR requires at most three calls per productive refinement iteration.
The accepted sub-criteria are then re-judged together by the pairwise criterion judge.

Table~\ref{tab:btcr_efficiency} reports empirical call savings on the Split-C validation set.
Compared with the plain per-criterion loop, BTCR reduces Stage-2.5 refinement calls from $96{,}571$ to $10{,}908$, an $88.7\%$ reduction, or $8.9\times$ fewer calls.
The largest savings come from redundancy filtering, which would otherwise require one LLM call for each generated sub-criterion.
The reported plain-loop decomposition count is conservative because it is estimated from the initial tie count and does not include additional tied criteria that may arise after intermediate re-judging.

\begin{table}[t]
    \centering
    \caption{
    \textbf{LLM-call savings from Batched Tie-Driven Criterion Refinement (BTCR).}
    We compare a plain per-criterion refinement loop with our batched implementation on the Split-C validation set.
    BTCR reduces Stage-2.5 refinement calls by $88.7\%$.
    }
    \label{tab:btcr_efficiency}
    \scalebox{0.9}{
    \begin{tabular}{L{4.0cm}|C{1.5cm}C{1.5cm}C{1.5cm}}
        \toprule
        \textbf{Phase} & \textbf{Plain loop} & \textbf{BTCR} & \textbf{Reduction} \\
        \midrule
        Decomposition & $16{,}795$ & $3{,}963$ & $-12{,}832$ \\
        Redundancy filter & $62{,}218$ & $3{,}953$ & $-58{,}265$ \\
        Conflict filter & $17{,}558$ & $2{,}992$ & $-14{,}566$ \\
        \midrule
        Total Stage-2.5 calls & $96{,}571$ & $10{,}908$ & $-85{,}663$ \\
        \bottomrule
    \end{tabular}
    }
\end{table}

\subsection{Additional WebDevJudge Transfer Results}
\label{app:webdev_transfer}

We additionally evaluate whether HPAG synthesized from BigCodeReward can be transferred to WebDevJudge~\cite{webdevjudge2025} without WebDevJudge-specific guidance synthesis. 
Unlike our main experiments, which focus on BigCodeReward, this setting tests a cross-dataset transfer scenario where the target benchmark is narrower and more focused on web-development preferences. 
Table~\ref{tab:webdev-transfer} reports the results.

\begin{table}[t]
    \centering
    \caption{
    \textbf{Additional WebDevJudge transfer results.}
    HPAG is synthesized from BigCodeReward and applied without WebDevJudge-specific guidance synthesis.
    Numbers are accuracy in percent; \textit{imp.} denotes improvement over the corresponding baseline.
    }
    \label{tab:webdev-transfer}
    \scalebox{0.9}{
    \begin{tabular}{L{4.2cm}|C{1.2cm}C{1.2cm}}
        \toprule
        \textbf{Method} & \textbf{Acc.} & \textbf{\textit{imp.}} \\
        \midrule
        Monolithic & $65.0$ & - \\
        Monolithic + HPAG & $67.7$ & $+2.7$ \\
        \midrule
        Pointwise RRD & $61.1$ & - \\
        Pointwise RRD + HPAG & $64.2$ & $+3.1$ \\
        \midrule
        Ours w/o HPAG & $68.0$ & - \\
        Ours w/ HPAG & $68.7$ & $+0.7$ \\
        \bottomrule
    \end{tabular}
    }
\end{table}

HPAG improves the monolithic judge from $65.0\%$ to $67.7\%$ and improves pointwise RRD from $61.1\%$ to $64.2\%$, suggesting that the synthesized guidance can provide useful preference signals beyond the source dataset. 
For our full pairwise criterion pipeline, HPAG gives a smaller gain, improving accuracy from $68.0\%$ to $68.7\%$. 
This indicates that the pairwise criterion pipeline itself already transfers reasonably well to WebDevJudge, while HPAG provides an additional but more modest improvement in this cross-dataset setting.

\section{Qualitative Case Study: Criterion-Level Evidence for Code Preference}
\label{app:case_rule30}

We provide a qualitative example to illustrate how \textsc{CriterAlign} differs from a monolithic LLM judge. 
The instruction asks the model to write a C++ program that visualizes the Rule 30 cellular automaton and prints each generation as ASCII output:

\begin{quote}
\emph{Write a C++ program that visualizes cellular automata (Rule 30) and prints each generation as ASCII output.}
\end{quote}

\begin{figure}
    \centering
    \includegraphics[width=0.8\linewidth]{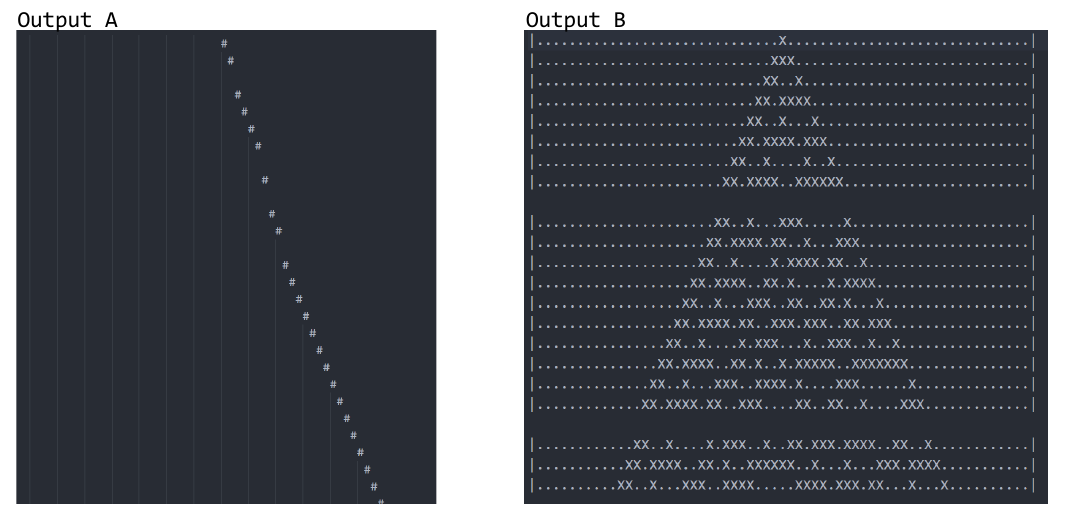}
    \caption{Caption}
    \label{fig:output}
\end{figure}

The pair consists of two candidate solutions, where Solution A is generated by Gemini-2.0-Flash and Solution B is generated by o1. 
The monolithic Qwen2.5-VL-32B judge selects Solution A, while \textsc{CriterAlign} selects Solution B. The output is shown in Figure~\ref{fig:output}.

\paragraph{Candidate solutions.}
Listing~\ref{lst:rule30_a} and Listing~\ref{lst:rule30_b} show the two candidate implementations. 
For brevity, we keep the original code structure and comments, since the differences in boundary handling and output formatting are directly relevant to the judgment.

\begin{lstlisting}[style=promptstyle,caption={Solution A for the Rule 30 cellular automaton task.},label={lst:rule30_a}]
#include <iostream>
#include <vector>

using namespace std;

// Function to calculate the next state of a cell based on Rule 30
bool rule30(bool left, bool center, bool right) {
    if (left && center && right) return false;
    if (left && center && !right) return false;
    if (left && !center && right) return true;
    if (left && !center && !right) return true;
    if (!left && center && right) return true;
    if (!left && center && !right) return false;
    if (!left && !center && right) return false;
    if (!left && !center && !right) return false;
    return false;
}

int main() {
    int num_generations = 40;
    int num_cells = 81;

    vector<bool> cells(num_cells, false);
    cells[num_cells / 2] = true;

    cout << endl << "Rule 30 Cellular Automaton:" << endl << endl;

    for (int i = 0; i < num_generations; ++i) {
        for (bool cell : cells) {
            cout << (cell ? '#' : ' ');
        }
        cout << endl;

        vector<bool> next_generation(num_cells);
        for (int j = 0; j < num_cells; ++j) {
            bool left = (j > 0) ? cells[j - 1] : false;
            bool center = cells[j];
            bool right = (j < num_cells - 1) ? cells[j + 1] : false;

            next_generation[j] = rule30(left, center, right);
        }

        cells = next_generation;
    }

    cout << endl;
    return 0;
}
\end{lstlisting}

\begin{lstlisting}[style=promptstyle,caption={Solution B for the Rule 30 cellular automaton task.},label={lst:rule30_b}]
#include <iostream>
#include <vector>
#include <string>
using namespace std;

const string RESET = "\x1b[0m";
const string BRIGHT_BLACK = "\x1b[90m";
const string BRIGHT_GREEN = "\x1b[92m";

int rule30(int left, int center, int right) {
    int neighborhood = (left << 2) | (center << 1) | right;
    switch (neighborhood) {
        case 4:
        case 3:
        case 2:
        case 1:
            return 1;
        default:
            return 0;
    }
}

int main() {
    const int width = 61;
    const int generations = 20;

    vector<int> currentRow(width, 0);
    currentRow[width / 2] = 1;

    cout << "|" << BRIGHT_BLACK;
    for (int cell : currentRow) {
        if (cell == 1) {
            cout << BRIGHT_GREEN << "X" << BRIGHT_BLACK;
        } else {
            cout << ".";
        }
    }
    cout << RESET << "|" << endl;

    for (int g = 0; g < generations; g++) {
        vector<int> nextRow(width, 0);
        for (int i = 0; i < width; i++) {
            int left   = currentRow[(i - 1 + width) % width];
            int center = currentRow[i];
            int right  = currentRow[(i + 1) % width];
            nextRow[i] = rule30(left, center, right);
        }

        cout << "|" << BRIGHT_BLACK;
        for (int cell : nextRow) {
            if (cell == 1) {
                cout << BRIGHT_GREEN << "X" << BRIGHT_BLACK;
            } else {
                cout << ".";
            }
        }
        cout << RESET << "|" << endl;

        currentRow = nextRow;
    }

    return 0;
}
\end{lstlisting}

\paragraph{Monolithic judgment.}
The monolithic baseline favors Solution A because it considers A to be simpler and more directly aligned with the request for ASCII output. 
Its rationale emphasizes that Solution A uses a straightforward representation of live cells with ``\#'' and dead cells with spaces, avoids additional formatting such as ANSI color codes, and has a clean implementation structure. 
In contrast, it penalizes Solution B for introducing ANSI escape codes and circular boundary handling, which the monolithic judge views as unnecessary complexity.

However, this judgment focuses heavily on surface-level simplicity and underweights several implementation and presentation details that can matter to human users when evaluating a cellular automaton visualization. 
In particular, the monolithic judge treats colored output as a deviation from the instruction, even though the output is still ASCII-based and remains readable in a terminal. 
It also treats boundary handling as an optional implementation detail rather than a meaningful design choice for a cellular automaton simulation.

\paragraph{\textsc{CriterAlign} judgment.}
In contrast, \textsc{CriterAlign} decomposes the comparison into multiple task-specific criteria and judges each criterion pairwise. 
The final prediction favors Solution B. 
The criterion-level evidence highlights three major advantages of Solution B.

First, Solution B provides a more visually interpretable terminal output by using ANSI escape codes to distinguish live and dead cells. 
This is reflected in criteria related to visual appeal and ease of interpretation. 
Second, Solution B handles boundary cells consistently using modular arithmetic, which gives a wrap-around grid and avoids special-casing the edges as permanently dead. 
This is reflected in correctness-oriented criteria about applying Rule 30 consistently across all cells. 
Third, Solution B maintains a fixed-width representation, making the visualization more stable when the number of cells or generations increases.

Table~\ref{tab:case_rule30_criteria} summarizes the criterion-level evidence produced by \textsc{CriterAlign}. 
The example shows that criterion-level decomposition can make the source of the final preference more transparent: rather than directly producing a single overall rationale, \textsc{CriterAlign} exposes which concrete aspects support Solution B.

\begin{table}[t]
\centering
\small
\caption{
Qualitative case study on a Rule 30 cellular automaton task. 
The monolithic baseline selects Solution A, mainly because it is simpler and avoids ANSI formatting. 
\textsc{CriterAlign} selects Solution B by aggregating criterion-level evidence related to visual interpretability, consistent boundary handling, and output robustness.
}
\label{tab:case_rule30_criteria}
\scalebox{0.8}{
\begin{tabular}{p{0.08\linewidth} p{0.48\linewidth} p{0.10\linewidth} p{0.16\linewidth} p{0.10\linewidth}}
\toprule
\textbf{ID} & \textbf{Criterion} & \textbf{Verdict} & \textbf{Mapped Aspect} & \textbf{Conf.} \\
\midrule
c11 
& Uses ANSI escape codes effectively to add color to the output, if implemented. 
& B 
& UI/UX design 
& High \\

c20 
& Ensures that the output is visually appealing and easy to interpret, even for large numbers of generations. 
& B 
& UI/UX design 
& High \\

t24 
& Ensures that the output remains consistent when the number of generations or cells is increased beyond the current configuration. 
& B 
& Correctness 
& High \\

t27 
& Provides a mechanism to dynamically adjust the number of generations or cells at runtime, rather than relying on hardcoded values. 
& B 
& Others 
& High \\

t38 
& Ensures that the Rule 30 logic is applied consistently across all cells, including those near the boundaries, without introducing hidden randomness or variability. 
& B 
& Correctness 
& High \\
\bottomrule
\end{tabular}
}
\end{table}

\paragraph{Discussion.}
This example illustrates both the strength and the diagnostic value of criterion-centric judging. 
The monolithic judge produces a plausible but coarse rationale: it prefers the simpler implementation and interprets ANSI color formatting as unnecessary. 
By contrast, \textsc{CriterAlign} surfaces more fine-grained evidence, including visual interpretability and boundary consistency, which are easy to overlook in a single-pass holistic judgment.

At the same time, the example also reveals why criterion filtering and evidence grounding are important. 
For instance, criterion t27 asks whether the number of generations or cells can be adjusted dynamically at runtime. 
Neither solution fully implements such a mechanism, yet the criterion judge still assigns the advantage to Solution B based on perceived extensibility. 
This type of weakly supported criterion is precisely the kind of noisy evidence that motivates our refinement and filtering components. 
Overall, the case demonstrates that \textsc{CriterAlign} does not merely change the final answer; it also provides a more inspectable reasoning trace that helps identify which criteria drive the preference and which criteria may require filtering.

\section{Limitations}
\label{app:limitations}

\textsc{CriterAlign} introduces additional inference cost compared with a monolithic LLM judge because it performs criterion generation, criterion-level pairwise judging, and optional refinement and swap-consistency filtering. 
This cost is a trade-off for more structured, interpretable, and human-aligned preference prediction; highly latency-sensitive evaluation settings may benefit from a distilled or selectively triggered version of the pipeline.

Our evaluation focuses on BigCodeReward~\citep{bigcodearena2025} and WebDevJudge~\citep{webdevjudge2025}. 
While these benchmarks cover diverse code-preference scenarios, human preferences may vary across domains, user expertise, deployment contexts, and available evidence, such as hidden tests, UI behavior, maintainability, security, or explanation quality. 
Future work can extend HPAG synthesis to additional code-preference datasets and specialized domains.

HPAG is injected as natural-language guidance into frozen judge models, making it broadly applicable but still dependent on the judge's instruction-following ability and context budget. 
Compressing the guidance or distilling \textsc{CriterAlign} into a more efficient judge are promising directions.

\section{Broader Impact}
\label{app:broader_impact}

This work aims to improve the transparency and reliability of automated evaluation for code-generation systems. 
Compared with monolithic preference prediction, \textsc{CriterAlign} exposes criterion-level evidence for why one response is preferred over another, which can help researchers diagnose model failures and compare systems more carefully.

The framework may also support more interpretable model selection and reward modeling by providing richer feedback than a single scalar score or binary preference label. 
For example, criterion-level verdicts can help identify whether a preference is driven by correctness, clarity, efficiency, maintainability, visual quality, or other task-dependent factors.

As with any automated judge, \textsc{CriterAlign} should be used with human oversight, especially in settings where correctness, safety, or security matters. 
It is intended as a tool for research evaluation and diagnostic analysis, not as a replacement for execution-based testing, expert review, or security validation. 
When used for model optimization or deployment decisions, automated judgments should be validated against human preferences and task-specific tests.

\input{sec/appendix_prompts_ascii}

\input{sec/appendix_hpag}

%% file: tables/point2pair.tex
\begin{table}[t]
    \caption{
    \textbf{From pointwise rubrics to pairwise preference judging.}
    Rows differ in how criteria are generated, refined, judged, and aggregated.
    \textbf{Criterion Judge} denotes the per-criterion decision mechanism: a \emph{pointwise judge} scores responses A and B independently with binary YES/NO decisions, while a \emph{pairwise judge} compares both responses directly and outputs A/B/tie for each criterion.
    \textbf{Aggregator} reduces criterion-level outputs to a final preference, either through weighted score comparison or through a final pairwise LLM judge.
    \textbf{Prompt Type} distinguishes the original general-purpose phrasing from BigCodeArena-style code-specific phrasing.
    \criteriongen denotes criterion generation; \textbf{RD} and \textbf{RF} denote RRD's recursive decomposition and filtering; \textbf{BTCR} denotes our batched tie-driven criterion refinement; \textbf{SCF} denotes swap-consistency criterion filtering; \textbf{HPAG} denotes Human-Preference-Aligned Guidance.
    Numbers are accuracy in percent.
    }
    \label{tab:point2pair}
    \centering
    \scalebox{0.65}{
    \begin{tabular}{L{3.2cm}|C{3.5cm}C{3.9cm}C{4.3cm}|C{0.8cm}C{0.8cm}|C{1.0cm}}
    
        \toprule
        \multirow{2}{*}{\textbf{Family}}
        & \multirow{2}{*}{\textbf{Variant}}
        & \multirow{2}{*}{\textbf{Criterion Judge}}
        & \multirow{2}{*}{\textbf{Aggregator}}
        & \multicolumn{2}{c|}{\textbf{Prompt Type}}
        & \multirow{2}{*}{\textbf{Acc.}} \\
        & & & & General & Code & \\
        \midrule

        Baseline
        & Monolithic
        & -
        & pairwise LLM judge
        & 
        & \checkmark
        & 60.4 \\

        \midrule
        \multirow{3}{*}{LLM Rubrics}
        & w/ \criteriongen 
        & pointwise judge
        & uniform weighted sum
        & \checkmark
        & 
        & 47.5 \\

        & w/ \criteriongen
        & pointwise judge
        & LLM-assigned weighted sum
        & \checkmark
        & 
        & 49.6 \\

        & w/ \criteriongen
        & pointwise judge
        & pairwise LLM judge
        & \checkmark
        & 
        & 50.3 \\

        \midrule
        \multirow{6}{*}{Pointwise RRD}
        & w/ \criteriongen, RD, RF
        & pointwise judge
        & uniform weighted sum
        & \checkmark
        & 
        & 51.6 \\

        & w/ \criteriongen, RD, RF
        & pointwise judge
        & LLM-assigned weighted sum
        & \checkmark
        & 
        & 55.0 \\

        & w/ \criteriongen, RD, RF
        & pointwise judge
        & pairwise LLM judge
        & \checkmark
        & 
        & 53.3 \\

        & w/ \criteriongen, RD, RF
        & pointwise judge
        & uniform weighted sum
        & 
        & \checkmark
        & 51.7 \\

        & w/ \criteriongen, RD, RF
        & pointwise judge
        & LLM-assigned weighted sum
        & 
        & \checkmark
        & 54.4 \\

        & w/ \criteriongen, RD, RF
        & pointwise judge
        & pairwise LLM judge
        & 
        & \checkmark
        & 51.2 \\

        \midrule
        Pairwise RRD 
        & w/ \criteriongen, RD, RF
        & pairwise judge
        & pairwise LLM judge
        & 
        & \checkmark
        & 61.8 \\

        \midrule
        \textbf{\textsc{CriterAlign}}
        & w/ \criteriongen, BTCR, SCF
        & pairwise judge w/ HPAG
        & pairwise LLM judge w/ HPAG
        & 
        & \checkmark
        & $\mathbf{66.3}$ \\

        \bottomrule

    \end{tabular}
    }
\end{table}

%% file: sec/appendix_prompts_ascii.tex

\section{Prompt Templates}
\label{app:prompts}

This appendix lists the prompt templates used in \textsc{CriterAlign}. We use bracketed placeholders such as \texttt{\{INSTRUCTION\}}, \texttt{\{ANSWER\_A\}}, and \texttt{\{GUIDANCE\}} to denote fields filled at inference time. Swap-consistency criterion filtering does not require a separate prompt: it reuses the pairwise criterion-judging prompt with the candidate responses and their artifacts swapped.

\subsection{Human rationale reconstruction}
\label{app:prompt-human-rationale-enhanced}

\begin{lstlisting}[style=promptstyle,caption={Human rationale reconstruction prompt template.}]
You are a human preference analyst. A human evaluator compared two code solutions (A and B) for a programming task and provided both per-aspect judgments and an overall preference. Your job is to construct the most likely reasoning behind the human's overall choice, grounded in their aspect-level judgments.

You are given:
1. The user's programming instruction
2. Both solutions (code + execution output + screenshots if available)
3. The human's per-aspect votes (correctness, efficiency, explainability, maintainability, UI/UX design)
4. The human's overall vote

Your task: Synthesize a coherent rationale that explains the human's overall preference, grounded in their aspect-level judgments. Where aspects disagree with the overall vote, explain the likely reasoning for why certain aspects were weighted more heavily. Think like a practical developer, not a formal reviewer.

**Human Aspect Votes:**
{HUMAN_ASPECT_VOTES}

**Human Overall Vote:** {HUMAN_VOTE_LABEL}

**Input**
<|Instruction|>
{INSTRUCTION}

<|The Start of Assistant A's Answer|>
{ANSWER_A}{SCREENSHOT_A_SECTION}{VISUAL_A_SECTION}<|The End of Assistant A's Answer|>

<|The Start of Assistant B's Answer|>
{ANSWER_B}{SCREENSHOT_B_SECTION}{VISUAL_B_SECTION}<|The End of Assistant B's Answer|>

**Output Format**
Return exactly one JSON object. The "reasoning" field should be a single paragraph of 100-180 words providing an aspect-grounded explanation of the human's preference. Any quotation marks within the text should be properly escaped for valid JSON.

```json
{
  "reasoning": "...",
  "key_factors": ["factor1", "factor2", "factor3"],
  "primary_aspect": "correctness"|"efficiency"|"explainability"|"maintainability"|"ui_ux_design"|"visual_output"|"completeness",
  "aspect_analysis": {
    "correctness": {"vote": "A"|"B"|"Tie"|null, "importance": "high"|"medium"|"low"|"not_applicable", "explanation": "..."},
    "efficiency": {"vote": "A"|"B"|"Tie"|null, "importance": "high"|"medium"|"low"|"not_applicable", "explanation": "..."},
    "explainability": {"vote": "A"|"B"|"Tie"|null, "importance": "high"|"medium"|"low"|"not_applicable", "explanation": "..."},
    "maintainability": {"vote": "A"|"B"|"Tie"|null, "importance": "high"|"medium"|"low"|"not_applicable", "explanation": "..."},
    "ui_ux_design": {"vote": "A"|"B"|"Tie"|null, "importance": "high"|"medium"|"low"|"not_applicable", "explanation": "..."}
  },
  "decisive_aspects": ["correctness", "ui_ux_design"]
}
```

Requirements:
- "reasoning": A single paragraph (100-180 words) explaining the human's likely reasoning, grounded in the aspect votes
- "key_factors": 2-5 short phrases identifying the most decisive factors
- "primary_aspect": The single most important aspect that drove the overall preference
- "aspect_analysis": For each of the five aspects, echo back the human vote, rate its importance to the overall decision, and briefly explain why
- "decisive_aspects": List the 1-3 aspects that most strongly influenced the overall preference
\end{lstlisting}

\subsection{Human-preference-aligned guidance synthesis}
\label{app:prompt-guidance-synthesis-v2}

\begin{lstlisting}[style=promptstyle,caption={Human-preference-aligned guidance synthesis prompt template.}]
You are a meta-analyst for an LLM-as-a-judge system that predicts human pairwise preferences on code-generation tasks. Your job is to analyze how human evaluators reason about code quality compared to how the LLM judge reasons, and produce concrete **prompt guidance** -- both at the global level and for each task category -- to make the LLM judge more human-aligned.

## Background

We have a pipeline that evaluates pairs of code solutions (A vs B) for programming tasks. It works in three stages:
1. **Criterion generation**: An LLM generates 16-20 atomic evaluation criteria for the given task
2. **Per-criterion judging**: For each criterion, the LLM judges which solution is better (A/B/tie)
3. **Final judging**: The LLM makes an overall preference decision using the criterion judgments as evidence

At runtime, a sample's task category is known. Each of the three stage-LLMs will be shown the **global guidance for its stage concatenated with the category-specific guidance for its stage**. So for every category we need the same four kinds of guidance as the global block, each tuned to that category's failure modes -- not a short add-on.

The LLM judge currently achieves ~60% agreement with human evaluators. We want to close this gap.

We generated "reverse-engineered human rationales" -- for each sample, we showed the LLM the same code pair but told it the human's actual vote and asked it to explain WHY the human likely chose that preference. We also have the LLM judge's original rationale for each sample.

## Task categories
The samples span six categories: Web Development, Game Development, Creative Coding, Diagram Creation, Scientific Computing, and Problem Solving.

## Your input

### Aggregate statistics
{AGGREGATE_STATS}

### Concrete examples

Below are representative samples showing human rationale vs LLM judge rationale. Pay special attention to **disagreement cases** (where the LLM judge got a different answer from the human).

{SAMPLE_CASES}

## Your task

Analyze the patterns of divergence between human and LLM reasoning -- both cross-cutting patterns that appear across all categories AND patterns that are distinctive to particular categories -- then produce concrete prompt guidance at two levels:

1. **Global guidance** (applies to every sample, cross-cutting patterns): same structure as before -- divergence patterns plus stage-specific guidance for criterion generation, per-criterion judging, and final judging.

2. **Per-category guidance** (applies only when the sample is of that category, category-distinctive patterns): for **each** of the six categories, produce the *same four kinds* of content as the global block -- divergence patterns, criterion-generation guidance, criterion-judging guidance, final-judging guidance -- at **comparable length** to the global version (not a 50-word add-on).

The category-specific guidance should focus on patterns *particular* to that category, not restate the global guidance. If a pattern genuinely applies everywhere, put it in the global block; if it only manifests clearly in one category, put it there.

Focus on **actionable, specific** instructions, not generic platitudes. Each piece of guidance should address a real pattern you observed in the data.

Output exactly one JSON object:

```json
{
  "key_divergence_patterns": [
    "Pattern 1: ...",
    "Pattern 2: ...",
    "Pattern 3: ..."
  ],
  "criterion_generation_guidance": "Concrete cross-cutting instructions to add to the criterion generation prompt. Focus on what kinds of criteria to generate and what to prioritize across all task types. 200-400 words.",
  "criterion_judging_guidance": "Concrete cross-cutting instructions to add to the per-criterion judging prompt. Focus on how to evaluate each criterion in a more human-aligned way across all task types. 200-400 words.",
  "final_judging_guidance": "Concrete cross-cutting instructions to add to the final overall judging prompt. Focus on how to weigh different factors when making the overall preference decision across all task types. 200-400 words.",
  "category_specific_guidance": {
    "web_development": {
      "key_divergence_patterns": [
        "Web-specific pattern 1: ...",
        "Web-specific pattern 2: ..."
      ],
      "criterion_generation_guidance": "Category-specific instructions for criterion generation on web tasks. Focus on what concrete properties matter most for web (visual layout, interactivity, responsiveness, etc.) and which categories of criteria (UI polish, state handling, accessibility) historically drive human preference here. Do NOT repeat global guidance; only say what is distinctive to web. 200-400 words.",
        "criterion_judging_guidance": "Category-specific instructions for per-criterion judging on web tasks. Focus on evidence patterns that reliably settle judgments in this category (e.g., how screenshots should be weighed, when render correctness trumps code quality). Do NOT repeat global guidance. 200-400 words.",
      "final_judging_guidance": "Category-specific instructions for the final overall judge on web tasks. Focus on how to weigh aspects when deciding overall preference for web -- which aspects tend to dominate, which tiebreakers humans use. Do NOT repeat global guidance. 200-400 words."
    },
    "game_development": {
      "key_divergence_patterns": ["...", "..."],
      "criterion_generation_guidance": "200-400 words, game-specific.",
      "criterion_judging_guidance": "200-400 words, game-specific.",
      "final_judging_guidance": "200-400 words, game-specific."
    },
    "creative_coding": {
      "key_divergence_patterns": ["...", "..."],
      "criterion_generation_guidance": "200-400 words, creative-specific.",
      "criterion_judging_guidance": "200-400 words, creative-specific.",
      "final_judging_guidance": "200-400 words, creative-specific."
    },
    "diagram_creation": {
      "key_divergence_patterns": ["...", "..."],
      "criterion_generation_guidance": "200-400 words, diagram-specific.",
      "criterion_judging_guidance": "200-400 words, diagram-specific.",
      "final_judging_guidance": "200-400 words, diagram-specific."
    },
    "scientific_computing": {
      "key_divergence_patterns": ["...", "..."],
      "criterion_generation_guidance": "200-400 words, scientific-specific.",
      "criterion_judging_guidance": "200-400 words, scientific-specific.",
      "final_judging_guidance": "200-400 words, scientific-specific."
    },
    "problem_solving": {
      "key_divergence_patterns": ["...", "..."],
      "criterion_generation_guidance": "200-400 words, problem-solving-specific.",
      "criterion_judging_guidance": "200-400 words, problem-solving-specific.",
      "final_judging_guidance": "200-400 words, problem-solving-specific."
    }
  }
}
```

Requirements:

- **Every category must have all four fields filled out at the stated length.** Short or missing category blocks are not acceptable -- the downstream pipeline assumes each stage can pull substantive category guidance.
- **Each guidance section should be self-contained instructions that can be directly injected into a prompt.** Do not write meta-commentary like "For web tasks, the model should consider..." -- write the guidance itself as if addressing the stage-LLM directly ("Generate criteria that capture visual responsiveness..." / "When judging this criterion, prefer the solution whose screenshot shows...").
- **Be specific**: reference concrete patterns (e.g., "humans prioritize visual correctness over code structure for web tasks when screenshots diverge") rather than generic advice (e.g., "consider the user's perspective").
- **Ground every recommendation in observed patterns from the data above.** If a category has too few samples in the input to support strong category-distinctive guidance, say so in its `key_divergence_patterns` and keep the stage guidance conservative, but still produce substantive text.
- **Avoid repetition.** Category guidance should focus on what's *distinctive* to that category. Cross-cutting advice belongs in the global block.
- **Focus on the patterns most likely to close the human-LLM agreement gap.**

Important operational note for category guidance:

The downstream pipeline will CONCATENATE global stage-guidance + category stage-guidance for each of the three LLMs. So the category guidance is read *in addition to* the global guidance -- not as a replacement. Design the category guidance to **complement and specialize** the global guidance without restating it. If a category's best practice conflicts with the global advice (e.g., global says "reward UI polish" but diagrams need render correctness instead), make that override explicit in the category block.

\end{lstlisting}

\subsection{Criterion generation}
\label{app:prompt-criterion-generation}

\begin{lstlisting}[style=promptstyle,caption={Criterion generation prompt template.}]
You are a code evaluation expert. Your task is to analyze a programming task and generate a set of atomic, task-specific evaluation criteria for comparing two candidate solutions (A and B).

Your goal is NOT to produce a generic checklist of everything that could be true of a solution. Instead, produce criteria that are most likely to determine overall human preference between the two responses.

Read the instruction and both solutions carefully. Generate criteria that are grounded in the actual task requirements and the concrete differences that could plausibly make a user prefer one solution over the other.

**Human-alignment guidance (derived from analysis of human-vs-LLM preference disagreements):**
{GUIDANCE}

**Category-specific guidance (for this task's category):**
{CATEGORY_GUIDANCE}

**Each criterion must be:**
1. **Atomic** -- one checkable claim, with no "and/or" combinations
2. **Specific** -- concrete observable properties, not abstractions like "good code"
3. **Judgeable** -- a downstream model can decide it from the code, answer text, and execution evidence
4. **Task-relevant** -- tied to the instruction's requirements, constraints, expected outputs, or user-valued qualities for this task
5. **Non-redundant** -- no near-duplicates or trivial rephrasings
6. **Comparative-useful** -- the criterion should capture a property on which the two responses could plausibly differ in a way that affects user preference

**Important goal: prefer preference-driving criteria over adequacy-only checks.**
- Prefer criteria that would actually help distinguish which solution is better.
- Avoid criteria that both solutions are likely to satisfy equally, unless a noticeable difference on that criterion would strongly affect overall preference.
- Avoid turning generic best practices into criteria unless they are genuinely relevant to this task and likely to matter to human preference.
- At most 2 criteria may be pure minimum-adequacy checks.
- At least half of the criteria should be clearly comparative / preference-driving.

**Symmetry / anti-bias requirement**
- Treat Solution A and Solution B symmetrically.
- Do NOT treat either response as the default, baseline, or reference answer.
- If the positions of A and B were swapped, the generated criteria should remain semantically identical.
- Phrase criteria in a response-neutral way such as "The solution ..." or "The response ...", not "Solution A ..." or "Whether B improves on A ...".

**Target:** 16-20 high-quality criteria. Prefer quality over quantity.

**Prioritization guidance**
When choosing what criteria to include, prioritize:
1. Whether the solution correctly fulfills the user's actual request
2. Whether it handles important edge cases or failure modes relevant to this task
3. Whether it respects explicit constraints in the instruction
4. Whether one response is meaningfully more useful, complete, robust, or user-aligned
5. Only then consider secondary qualities like readability, comments, or maintainability, and only if they are likely to affect user preference here

---

<|Instruction|>
{INSTRUCTION}

<|The Start of Assistant A's Answer|>
{ANSWER_A}{SCREENSHOT_A_SECTION}{VISUAL_A_SECTION}<|The End of Assistant A's Answer|>

<|The Start of Assistant B's Answer|>
{ANSWER_B}{SCREENSHOT_B_SECTION}{VISUAL_B_SECTION}<|The End of Assistant B's Answer|>

---

Return ONLY valid JSON matching this exact schema (no additional text, no markdown explanation):

```json
{
  "criteria": [
    {
      "id": "c1",
      "criterion": "The solution correctly handles the case where the input list is empty by returning an empty result.",
      "rationale": "The task description explicitly requires handling empty inputs, and a difference here would materially affect user preference.",
      "evidence_basis": ["instruction", "code", "execution_output"]
    }
  ]
}
```

Field definitions:
- `id`: unique string, sequential like "c1", "c2", "c3", ...
- `criterion`: a single atomic checkable statement about one specific property
- `rationale`: short explanation of why this criterion is relevant to this specific task
- `evidence_basis`: subset of ["instruction", "code", "execution_output", "screenshot"]

Generation process:
1. Identify the key requirements, explicit constraints, likely failure modes, and meaningful quality differences from the instruction and both solutions
2. Draft criteria that are grounded in those concrete requirements and likely to influence overall preference
3. Remove any criterion that is vague, redundant, weakly judgeable, or unlikely to distinguish the two responses
4. Remove excess adequacy-only checks if they dominate the list
5. Verify each surviving criterion is atomic, judgeable, response-neutral, and useful for pairwise comparison
6. Output JSON only

\end{lstlisting}

\subsection{Pairwise criterion judging}
\label{app:prompt-criterion-judge-batch}

\begin{lstlisting}[style=promptstyle,caption={Pairwise criterion judging prompt template.}]
You are a code evaluation judge. Your task is to compare two candidate solutions (A and B) against a specific list of evaluation criteria.

For each criterion, determine which solution better satisfies it based on the code implementations and execution results provided. Additionally, classify each criterion into the human evaluation aspect it most closely relates to.

**Human-alignment guidance (derived from analysis of human-vs-LLM preference disagreements):**
{GUIDANCE}

**Category-specific guidance (for this task's category):**
{CATEGORY_GUIDANCE}

**Judgment values:**
- `"A"` -- Solution A clearly better satisfies this criterion
- `"B"` -- Solution B clearly better satisfies this criterion
- `"tie"` -- Both solutions satisfy this criterion to a genuinely comparable degree
- `"insufficient_evidence"` -- Cannot determine from the available information

**Confidence:**
- `"high"` -- Clear, unambiguous evidence in the code or execution output
- `"medium"` -- Reasonable inference from available evidence
- `"low"` -- Limited evidence; judgment is uncertain

**Mapped aspect:**
Classify which human evaluation aspect this criterion most closely relates to:
- `"correctness"` -- whether the solution produces correct results, handles requirements properly
- `"efficiency"` -- performance, resource usage, algorithmic complexity
- `"explainability"` -- clarity of code logic, comments, documentation, readability
- `"maintainability"` -- code structure, modularity, extensibility, best practices
- `"ui_ux_design"` -- visual output quality, user interface, user experience, rendering
- `"OTHERS"` -- does not fit any of the above five aspects

**Core judging principles**
1. Judge each criterion **symmetrically**. Do not treat Solution A as the default or reference answer.
2. If the positions of A and B were swapped, the judgment should swap accordingly.
3. Base the judgment on **comparative evidence**, not merely on whether each solution individually clears a minimum bar.
4. Use `"tie"` only when the available evidence indicates the two solutions are genuinely comparable on this specific criterion.
5. Do NOT default to `"A"` when the difference is subtle. If the evidence slightly but meaningfully favors B, choose `"B"`.
6. If both solutions are flawed in different ways, still choose the one that better satisfies the criterion unless they are genuinely comparable.
7. Do not let your judgment on one criterion leak into another; evaluate each criterion independently.

**Anti-bias requirement**
- Be strictly position-invariant.
- Do not assume A is better because it appears first.
- Do not use style, verbosity, or answer order as a tiebreaker unless the criterion explicitly concerns those aspects.

---

<|Instruction|>
{INSTRUCTION}

<|The Start of Assistant A's Answer|>
{ANSWER_A}{SCREENSHOT_A_SECTION}{VISUAL_A_SECTION}<|The End of Assistant A's Answer|>

<|The Start of Assistant B's Answer|>
{ANSWER_B}{SCREENSHOT_B_SECTION}{VISUAL_B_SECTION}<|The End of Assistant B's Answer|>

---

**Criteria to evaluate:**

{CRITERIA_LIST}

---

Evaluate ALL criteria listed above. Return ONLY valid JSON matching this exact schema:

```json
{
  "criterion_results": [
    {
      "criterion_id": "c1",
      "criterion": "...",
      "judgment": "A",
      "confidence": "high",
      "rationale": "Solution A handles this correctly because ..., while Solution B ...",
      "evidence_basis": ["code"],
      "mapped_aspect": "correctness"
    }
  ]
}
```

Requirements:
- Include one entry per criterion in the same order as the input list
- `judgment` must be one of: "A", "B", "tie", "insufficient_evidence"
- `confidence` must be one of: "high", "medium", "low"
- `rationale` must cite specific comparative evidence from the code or execution output
- `evidence_basis` should reflect what you actually used: subset of ["instruction", "code", "execution_output", "screenshot"]
- `mapped_aspect` must be one of: "correctness", "efficiency", "explainability", "maintainability", "ui_ux_design", "OTHERS"
- Do not skip any criterion -- use "insufficient_evidence" if you cannot determine a clear winner
- Do not rewrite the criterion; evaluate it as written
- Keep each rationale focused on this criterion only
\end{lstlisting}

\subsection{Tie-driven criterion decomposition}
\label{app:prompt-criterion-decompose-batch}

\begin{lstlisting}[style=promptstyle,caption={Tie-driven criterion decomposition prompt template.}]
Role: You are a criterion designer for an LLM-as-judge code evaluation system.

Context: A pairwise judge compared two candidate solutions (A and B) against a set of evaluation criteria. The criteria listed below were each judged as a **tie** -- both solutions satisfy them to a comparable degree at the current level of granularity.

Your task: For EACH tied criterion, propose exactly TWO new, more granular sub-criteria derived from that parent criterion. These sub-criteria should target specific dimensions where Solution A and Solution B might plausibly differ, even though they appear tied at the parent level.

What "more granular" means (requirements):
- Each sub-criterion must target a specific, discriminative dimension of quality that the parent criterion does not distinguish (e.g., handling of a particular edge case, efficiency of a specific operation, clarity of a specific code section).
- Sub-criteria should collectively cover the parent criterion's intent -- do not lose critical information.
- Each sub-criterion must be consistently judgeable from the code, execution output, or answer text.
- Each sub-criterion must be task-specific (tied to the instruction's requirements), not generic advice.
- Each sub-criterion should be written as a single atomic criterion -- one checkable claim, no "and/or".
- Sub-criteria MUST NOT repeat or overlap with the other existing criteria listed below.
- Sub-criteria MUST NOT reproduce or quote the candidate solutions' code.

Tips for effective decomposition:
- Look at HOW the two solutions satisfy the parent criterion differently -- decompose along that axis.
- Prefer criteria that test observable, concrete properties (output correctness, specific API usage, error handling for a particular case) over subjective qualities.
- If the parent criterion is about correctness, decompose into specific test cases or edge cases.
- If the parent criterion is about code quality, decompose into specific measurable aspects (e.g., function decomposition, naming, duplication).

---

<|Instruction|>
{INSTRUCTION}

<|The Start of Assistant A's Answer|>
{ANSWER_A}{SCREENSHOT_A_SECTION}{VISUAL_A_SECTION}<|The End of Assistant A's Answer|>

<|The Start of Assistant B's Answer|>
{ANSWER_B}{SCREENSHOT_B_SECTION}{VISUAL_B_SECTION}<|The End of Assistant B's Answer|>

<|Tied Criteria (each needs decomposition into 2 sub-criteria)|>
{TIED_CRITERIA}

<|Other Existing Criteria (new sub-criteria must NOT overlap with these)|>
{OTHER_CRITERIA}

---

Return ONLY valid JSON matching this exact schema:

```json
{
  "decompositions": [
    {
      "parent_id": "c3",
      "sub_criteria": [
        {
          "criterion": "The solution correctly handles the edge case where the input list is empty by returning an appropriate default value.",
          "rationale": "The parent criterion assessed general correctness equally, but this sub-criterion targets a specific edge case where the solutions may differ.",
          "evidence_basis": ["code", "execution_output"]
        },
        {
          "criterion": "The solution avoids redundant iterations over the input data by using a single-pass algorithm.",
          "rationale": "The parent criterion assessed efficiency equally, but this sub-criterion targets a specific algorithmic choice where the solutions may differ.",
          "evidence_basis": ["code"]
        }
      ]
    }
  ]
}
```

Requirements:
- Include one entry per tied criterion, using the same `parent_id` from the input
- Each entry must have exactly 2 sub-criteria
- Sub-criteria must be specific to the parent criterion's domain, not generic
- Do not skip any tied criterion
\end{lstlisting}

\subsection{Batch redundancy filtering}
\label{app:prompt-rubric-redundancy-batch}

\begin{lstlisting}[style=promptstyle,caption={Batch redundancy filtering prompt template.}]
You are checking whether each new rubric substantially overlaps with ANY of the existing rubrics. For each new rubric, output whether it is redundant or not.

Definition of substantial overlapping:
- The new rubric has the same intent as an existing rubric, or is a strict subset/superset of it, or >= 70% of its meaning is covered by the existing rubric so that applying both would not materially change scoring outcomes.
- Match on meaning, not wording. Treat synonyms, paraphrases, and inverses with the same effect as overlapping (e.g., "be concise" ~= "avoid unnecessary verbosity").
- Ignore trivial phrasing, tone, and example differences unless they change the requirement.

Edge cases:
- If scopes are disjoint (different capability/goal) -> not redundant.
- If the new rubric adds only minor qualifiers (e.g., "clearly"/"appropriately") without changing evaluation -> redundant.
- If the new rubric merely narrows the context while keeping the same criterion (subset) or broadens it (superset) -> redundant.

---

Input format:
<|Existing Accepted Rubrics|>
{EXISTING_RUBRICS}

<|New Rubrics to Check|>
{NEW_RUBRICS}

---

For EACH new rubric, determine if it is redundant with any existing rubric. Return ONLY valid JSON matching this exact schema. No other text is allowed:

```json
{
  "results": [
    {"id": "t1", "redundant": true},
    {"id": "t2", "redundant": false}
  ]
}
```

Requirements:
- Include one entry per new rubric, in the same order as the input
- Use the same `id` from the input
\end{lstlisting}

\subsection{Batch conflict filtering}
\label{app:prompt-rubric-conflicting-batch}

\begin{lstlisting}[style=promptstyle,caption={Batch conflict filtering prompt template.}]
You are checking whether each new rubric expresses opposite meaning of ANY of the existing rubrics. For each new rubric, output whether it is conflicting or not.

Definition of opposition:
- "Opposite" means the new rubric asserts the negation or reverse polarity of the same requirement, property, or direction as an existing rubric.
- Examples:
    - require X <-> forbid/avoid X
    - must include X <-> must NOT include X
    - prefer more of X <-> prefer less of X (same X, opposite direction)
    - answer should be optimistic <-> answer should be pessimistic
- Do NOT flag different targets or contexts.
- Do NOT flag orthogonal dimensions (e.g., tone vs citations or "be clear" vs "be concise")
- Do NOT flag mere differences in emphasis, strength, scope, or style.
- Do NOT flag stricter/looser thresholds unless they clearly reverse direction on the same axis (e.g., "maximize brevity" vs "maximize elaboration" = opposite; "<= 120 words" vs "<= 150 words" = NOT opposite).

---

Input format:
<|Existing Accepted Rubrics|>
{EXISTING_RUBRICS}

<|New Rubrics to Check|>
{NEW_RUBRICS}

---

For EACH new rubric, determine if it is conflicting with any existing rubric. Return ONLY valid JSON matching this exact schema. No other text is allowed:

```json
{
  "results": [
    {"id": "t1", "conflicting": false},
    {"id": "t2", "conflicting": false}
  ]
}
```

Requirements:
- Include one entry per new rubric, in the same order as the input
- Use the same `id` from the input
\end{lstlisting}

\subsection{Final judge with criterion evidence}
\label{app:prompt-judge-prompt-with-criteria}

\begin{lstlisting}[style=promptstyle,caption={Final judge with criterion evidence prompt template.}]
You are a code-review judge assigned to compare two candidate solutions (A and B) against a user's programming request. Your job is to evaluate each submission and choose an overall winner based on how well each solution fulfills the user's actual request.

**Primary objective**
Your primary goal is to determine which solution a human evaluator would prefer overall, with the highest weight placed on:
1. Correctly implementing the requested functionality
2. Respecting explicit constraints and requirements
3. Avoiding important errors, omissions, or misleading behavior

You may also consider efficiency, explainability, maintainability, and UI/UX when relevant, but these should not outweigh major correctness or requirement-fulfillment differences.

**Human-alignment guidance (derived from analysis of human-vs-LLM preference disagreements):**
{GUIDANCE}

**Category-specific guidance (for this task's category):**
{CATEGORY_GUIDANCE}

**Winner Options**
- `"A"`: Solution A is clearly better overall
- `"B"`: Solution B is clearly better overall
- `"Tie"`: Both solutions are genuinely comparable overall

**Overall evaluation principles**
1. Evaluate the two responses **symmetrically**. Do not treat Solution A as the default or reference.
2. Focus on the **most decisive differences**, not on counting superficial advantages.
3. Do not mechanically follow the majority of per-criterion labels. Many local ties or minor advantages do not necessarily imply an overall tie.
4. A small number of high-impact differences may outweigh many minor equalities.
5. Use `"Tie"` only when the solutions are genuinely comparable in the aspects that matter most to the user's request.
6. If the positions of A and B were swapped, your overall judgment should swap accordingly.
7. Do not favor A because it appears first, is longer, sounds more confident, or is more stylistically polished unless those qualities materially improve fulfillment of the user's request.

**Input Format**
<|Instruction|>
{INSTRUCTION}

<|The Start of Assistant A's Answer|>
{ANSWER_A}{SCREENSHOT_A_SECTION}{VISUAL_A_SECTION}<|The End of Assistant A's Answer|>

<|The Start of Assistant B's Answer|>
{ANSWER_B}{SCREENSHOT_B_SECTION}{VISUAL_B_SECTION}<|The End of Assistant B's Answer|>

**Per-Criterion Evaluation Results**
The following criterion-level judgments were produced by an independent evaluation step. Use them as additional evidence, but apply your own judgment for the overall preference.

{CRITERION_RESULTS}

**Output Format**
Return exactly one JSON object with this schema below. "reasoning" is a single paragraph explanation without line breaks. Any quotation marks within the text should be properly escaped for a valid JSON format.
```json
{
 "Overall": {
   "winner": "A"|"B"|"Tie",
   "reasoning": "..."
 }
}
```

Additional requirements for the reasoning:
- Explain the decisive factors behind the overall preference
- Emphasize major functional or requirement-related differences first
- Mention secondary factors only if they materially affected the outcome
- Do not justify the decision by saying one response came first or seemed like the default
\end{lstlisting}

%% file: sec/appendix_hpag.tex
\section{Synthesized Human-Preference-Aligned Guidance}
\label{app:hpag}

This appendix presents the synthesized Human-Preference-Aligned Guidance (HPAG) artifact used to steer the criterion generator, criterion judge, and final judge. The guidance was produced offline from training examples only, by comparing human preference rationales with the monolithic judge's rationales. It contains global divergence patterns and stage-specific guidance, followed by category-specific guidance for the six BigCodeReward task categories. The prompt used to synthesize this artifact is provided in Appendix~\ref{app:prompts}. The HPAG is synthesized by Claude Sonnet 4.6 for Qwen2.5-VL-32B.

\subsection{Global Human--LLM Divergence Patterns}
\label{app:hpag-global-patterns}

\begin{itemize}
    \item Pattern 1: Humans prioritize UI/UX design and visual correctness over code structure and efficiency when screenshots diverge from code.
    \item Pattern 2: Humans value completeness and adherence to requirements more than efficiency or maintainability when both solutions are correct.
    \item Pattern 3: Humans often weigh correctness and UI/UX design more heavily than maintainability and explainability, especially in interactive or visual tasks.
\end{itemize}

\subsection{Global Stage-Specific HPAG}
\label{app:hpag-global-guidance}

\paragraph{Criterion generator.}
When generating criteria, focus on properties that directly impact user experience and task-specific requirements. Prioritize correctness, UI/UX design, and visual correctness over abstract concepts like maintainability or efficiency unless explicitly requested. Ensure criteria are measurable and directly observable from the provided artifacts (code, screenshots, etc.). Include criteria for edge cases and error handling, as humans often value robustness and reliability. For tasks involving interactivity or visuals, generate criteria that assess responsiveness, usability, and visual appeal. Avoid overly technical criteria that are difficult to evaluate from the provided artifacts.

\paragraph{Criterion judge.}
When judging each criterion, prioritize evidence from screenshots and visual outputs over code structure or comments. For correctness, verify that the solution meets all requirements and handles edge cases. For UI/UX design, assess visual appeal, usability, and consistency. When evaluating maintainability, focus on code organization and modularity, but remember that humans often value correctness and user-facing features more. For efficiency, measure performance from the provided artifacts, but recognize that humans may overlook minor inefficiencies if the solution is correct and visually appealing. Always consider the task's primary goals and prioritize criteria that align with those goals.

\paragraph{Final judge.}
In the final judgment, weigh correctness and UI/UX design most heavily, especially if the solutions differ in these areas. Consider maintainability and explainability as secondary factors, but only if they significantly impact the primary goals. Use tiebreakers like completeness, visual appeal, and adherence to requirements to resolve close calls. Remember that humans often value solutions that are both correct and visually appealing, even if they are slightly less efficient or maintainable. When in doubt, lean towards the solution that better aligns with the task's primary objectives.

\subsection{Category-Specific HPAG}
\label{app:hpag-category-guidance}
The category-specific guidance is concatenated with the corresponding global stage guidance at inference time. We therefore write each block as additional instructions that specialize, rather than replace, the global guidance.

\subsubsection{Web Development}
\paragraph{Distinctive divergence patterns.}
\begin{itemize}
    \item Web-specific pattern 1: Humans prioritize visual correctness and interactivity over code structure and efficiency in web tasks.
    \item Web-specific pattern 2: Humans value responsiveness and accessibility as key factors in web development.
\end{itemize}
\paragraph{Criterion generator.}
For web tasks, generate criteria that focus on visual layout, interactivity, and responsiveness. Include specific criteria for visual correctness, such as alignment, spacing, and typography. Assess interactivity by evaluating how well the UI responds to user actions and whether it provides clear feedback. Measure responsiveness by checking how the solution adapts to different screen sizes and devices. Also, include criteria for accessibility, such as proper use of semantic HTML, ARIA attributes, and keyboard navigation support. Avoid overly technical criteria that are difficult to evaluate from screenshots alone.
\paragraph{Criterion judge.}
When judging web tasks, prioritize visual correctness and interactivity. Verify that the solution meets all UI/UX requirements and handles edge cases. For interactivity, assess how well the UI responds to user actions and whether it provides clear feedback. Measure responsiveness by checking how the solution adapts to different screen sizes and devices. Always consider accessibility features, such as proper use of semantic HTML, ARIA attributes, and keyboard navigation support. Use screenshots to evaluate visual correctness and interactivity, and rely on code snippets for assessing maintainability and efficiency only when necessary.
\paragraph{Final judge.}
In the final judgment for web tasks, weigh visual correctness and interactivity most heavily. Consider responsiveness and accessibility as secondary factors, but only if they significantly impact the primary goals. Use tiebreakers like completeness, visual appeal, and adherence to requirements to resolve close calls. Remember that humans often value solutions that are both visually appealing and interactive, even if they are slightly less efficient or maintainable. When in doubt, lean towards the solution that better aligns with the task's primary objectives.

\subsubsection{Game Development}
\paragraph{Distinctive divergence patterns.}
\begin{itemize}
    \item Game-specific pattern 1: Humans prioritize gameplay mechanics and visual appeal over code structure and efficiency in game tasks.
    \item Game-specific pattern 2: Humans value responsiveness and smooth animations as key factors in game development.
\end{itemize}
\paragraph{Criterion generator.}
For game tasks, generate criteria that focus on gameplay mechanics, visual appeal, and responsiveness. Include specific criteria for gameplay correctness, such as handling edge cases and ensuring smooth animations. Assess visual appeal by evaluating the overall design, color scheme, and consistency. Measure responsiveness by checking how well the game handles user input and maintains a smooth frame rate. Also, include criteria for smooth animations and transitions, as humans often value visual polish in games. Avoid overly technical criteria that are difficult to evaluate from screenshots alone.
\paragraph{Criterion judge.}
When judging game tasks, prioritize gameplay mechanics and visual appeal. Verify that the solution meets all gameplay requirements and handles edge cases. For visual appeal, assess the overall design, color scheme, and consistency. Measure responsiveness by checking how well the game handles user input and maintains a smooth frame rate. Always consider smooth animations and transitions, as humans often value visual polish in games. Use screenshots to evaluate gameplay mechanics and visual appeal, and rely on code snippets for assessing maintainability and efficiency only when necessary.
\paragraph{Final judge.}
In the final judgment for game tasks, weigh gameplay mechanics and visual appeal most heavily. Consider responsiveness and smooth animations as secondary factors, but only if they significantly impact the primary goals. Use tiebreakers like completeness, visual appeal, and adherence to requirements to resolve close calls. Remember that humans often value solutions that are both visually appealing and mechanically sound, even if they are slightly less efficient or maintainable. When in doubt, lean towards the solution that better aligns with the task's primary objectives.

\subsubsection{Creative Coding}
\paragraph{Distinctive divergence patterns.}
\begin{itemize}
    \item Creative-specific pattern 1: Humans prioritize visual correctness and interactivity over code structure and efficiency in creative tasks.
    \item Creative-specific pattern 2: Humans value artistic expression and innovation as key factors in creative coding.
\end{itemize}
\paragraph{Criterion generator.}
For creative tasks, generate criteria that focus on visual correctness, interactivity, and artistic expression. Include specific criteria for visual correctness, such as alignment, spacing, and typography. Assess interactivity by evaluating how well the solution responds to user actions and whether it provides clear feedback. Measure artistic expression by checking the uniqueness and creativity of the solution. Also, include criteria for innovation, such as novel approaches or unconventional techniques. Avoid overly technical criteria that are difficult to evaluate from screenshots alone.
\paragraph{Criterion judge.}
When judging creative tasks, prioritize visual correctness and interactivity. Verify that the solution meets all visual and interactive requirements and handles edge cases. For artistic expression, assess the uniqueness and creativity of the solution. Measure innovation by checking for novel approaches or unconventional techniques. Always consider artistic expression and innovation, as humans often value solutions that are both visually appealing and creatively unique. Use screenshots to evaluate visual correctness and interactivity, and rely on code snippets for assessing maintainability and efficiency only when necessary.
\paragraph{Final judge.}
In the final judgment for creative tasks, weigh visual correctness and artistic expression most heavily. Consider interactivity and innovation as secondary factors, but only if they significantly impact the primary goals. Use tiebreakers like completeness, visual appeal, and adherence to requirements to resolve close calls. Remember that humans often value solutions that are both visually appealing and creatively unique, even if they are slightly less efficient or maintainable. When in doubt, lean towards the solution that better aligns with the task's primary objectives.

\subsubsection{Diagram Creation}
\paragraph{Distinctive divergence patterns.}
\begin{itemize}
    \item Diagram-specific pattern 1: Humans prioritize visual correctness and clarity over code structure and efficiency in diagram tasks.
    \item Diagram-specific pattern 2: Humans value accuracy and completeness as key factors in diagram creation.
\end{itemize}
\paragraph{Criterion generator.}
For diagram tasks, generate criteria that focus on visual correctness, clarity, and accuracy. Include specific criteria for visual correctness, such as alignment, spacing, and typography. Assess clarity by evaluating the overall design, color scheme, and consistency. Measure accuracy by checking how well the solution handles edge cases and ensures smooth animations. Also, include criteria for completeness, as humans often value solutions that meet all requirements. Avoid overly technical criteria that are difficult to evaluate from screenshots alone.
\paragraph{Criterion judge.}
When judging diagram tasks, prioritize visual correctness and clarity. Verify that the solution meets all visual and interactive requirements and handles edge cases. For accuracy, assess the overall design, color scheme, and consistency. Measure completeness by checking how well the solution handles edge cases and ensures smooth animations. Always consider accuracy and completeness, as humans often value solutions that are both visually appealing and accurate. Use screenshots to evaluate visual correctness and clarity, and rely on code snippets for assessing maintainability and efficiency only when necessary.
\paragraph{Final judge.}
In the final judgment for diagram tasks, weigh visual correctness and accuracy most heavily. Consider clarity and completeness as secondary factors, but only if they significantly impact the primary goals. Use tiebreakers like completeness, visual appeal, and adherence to requirements to resolve close calls. Remember that humans often value solutions that are both visually appealing and accurate, even if they are slightly less efficient or maintainable. When in doubt, lean towards the solution that better aligns with the task's primary objectives.

\subsubsection{Scientific Computing}
\paragraph{Distinctive divergence patterns.}
\begin{itemize}
    \item Scientific-specific pattern 1: Humans prioritize correctness and efficiency over code structure and maintainability in scientific tasks.
    \item Scientific-specific pattern 2: Humans value reproducibility and documentation as key factors in scientific computing.
\end{itemize}
\paragraph{Criterion generator.}
For scientific tasks, generate criteria that focus on correctness, efficiency, and reproducibility. Include specific criteria for correctness, such as handling edge cases and ensuring accurate results. Assess efficiency by measuring performance from the provided artifacts. Measure reproducibility by checking whether the solution includes clear documentation and version control. Also, include criteria for documentation, as humans often value solutions that are well-documented. Avoid overly technical criteria that are difficult to evaluate from screenshots alone.
\paragraph{Criterion judge.}
When judging scientific tasks, prioritize correctness and efficiency. Verify that the solution meets all correctness requirements and handles edge cases. For efficiency, measure performance from the provided artifacts. Assess reproducibility by checking whether the solution includes clear documentation and version control. Always consider documentation, as humans often value solutions that are well-documented. Use screenshots to evaluate correctness and efficiency, and rely on code snippets for assessing maintainability and efficiency only when necessary.
\paragraph{Final judge.}
In the final judgment for scientific tasks, weigh correctness and efficiency most heavily. Consider reproducibility and documentation as secondary factors, but only if they significantly impact the primary goals. Use tiebreakers like completeness, visual appeal, and adherence to requirements to resolve close calls. Remember that humans often value solutions that are both correct and efficient, even if they are slightly less maintainable or documented. When in doubt, lean towards the solution that better aligns with the task's primary objectives.

\subsubsection{Problem Solving}
\paragraph{Distinctive divergence patterns.}
\begin{itemize}
    \item Problem-solving-specific pattern 1: Humans prioritize correctness and efficiency over code structure and maintainability in problem-solving tasks.
    \item Problem-solving-specific pattern 2: Humans value completeness and adherence to requirements as key factors in problem-solving.
\end{itemize}
\paragraph{Criterion generator.}
For problem-solving tasks, generate criteria that focus on correctness, efficiency, and completeness. Include specific criteria for correctness, such as handling edge cases and ensuring accurate results. Assess efficiency by measuring performance from the provided artifacts. Measure completeness by checking whether the solution meets all requirements. Also, include criteria for adherence to requirements, as humans often value solutions that are complete and accurate. Avoid overly technical criteria that are difficult to evaluate from screenshots alone.
\paragraph{Criterion judge.}
When judging problem-solving tasks, prioritize correctness and efficiency. Verify that the solution meets all correctness requirements and handles edge cases. For efficiency, measure performance from the provided artifacts. Assess completeness by checking whether the solution meets all requirements. Always consider adherence to requirements, as humans often value solutions that are complete and accurate. Use screenshots to evaluate correctness and efficiency, and rely on code snippets for assessing maintainability and efficiency only when necessary.
\paragraph{Final judge.}
In the final judgment for problem-solving tasks, weigh correctness and efficiency most heavily. Consider completeness and adherence to requirements as secondary factors, but only if they significantly impact the primary goals. Use tiebreakers like completeness, visual appeal, and adherence to requirements to resolve close calls. Remember that humans often value solutions that are both correct and efficient, even if they are slightly less maintainable or documented. When in doubt, lean towards the solution that better aligns with the task's primary objectives.

%% file: checklist.tex
\section*{NeurIPS Paper Checklist}

\begin{enumerate}

\item {\bf Claims}
    \item[] Question: Do the main claims made in the abstract and introduction accurately reflect the paper's contributions and scope?
    \item[] Answer: \answerYes{} 
    \item[] Justification:  The abstract and introduction clearly state that our contribution is a criterion-centric pairwise judging framework for code preference prediction, together with empirical evaluations on code preference benchmarks. We avoid claiming general human-alignment improvements beyond the evaluated datasets and judge settings.

\item {\bf Limitations}
    \item[] Question: Does the paper discuss the limitations of the work performed by the authors?
    \item[] Answer: \answerYes{} 
    \item[] Justification: We have discussed our limitations in Appendix~\ref{app:limitations}.

\item {\bf Theory assumptions and proofs}
    \item[] Question: For each theoretical result, does the paper provide the full set of assumptions and a complete (and correct) proof?
    \item[] Answer: \answerNA{} 
    \item[] Justification: The paper does not present theoretical results, formal theorems, or proofs; our contributions are methodological and empirical.

    \item {\bf Experimental result reproducibility}
    \item[] Question: Does the paper fully disclose all the information needed to reproduce the main experimental results of the paper to the extent that it affects the main claims and/or conclusions of the paper (regardless of whether the code and data are provided or not)?
    \item[] Answer: \answerYes{} 
    \item[] Justification: We report the implementation details in Section~\ref{sec:experiments}.

\item {\bf Open access to data and code}
    \item[] Question: Does the paper provide open access to the data and code, with sufficient instructions to faithfully reproduce the main experimental results, as described in supplemental material?
    \item[] Answer: \answerNo{} 
    \item[] Justification: The public datasets used in our experiments are accessible, and we provide detailed experimental protocols and prompts. We do not release the full code at submission time, but we include sufficient implementation details to reproduce the reported method. We will release the code after the paper gets accepted.

\item {\bf Experimental setting/details}
    \item[] Question: Does the paper specify all the training and test details (e.g., data splits, hyperparameters, how they were chosen, type of optimizer) necessary to understand the results?
    \item[] Answer: \answerYes{} 
    \item[] Justification: We report the implementation details in Section~\ref{sec:experiments}.

\item {\bf Experiment statistical significance}
    \item[] Question: Does the paper report error bars suitably and correctly defined or other appropriate information about the statistical significance of the experiments?
    \item[] Answer: \answerNo{} 
    \item[] Justification: We report deterministic accuracy on fixed benchmark splits, but do not include error bars for all experiments due to the high cost of repeated LLM judging.

\item {\bf Experiments compute resources}
    \item[] Question: For each experiment, does the paper provide sufficient information on the computer resources (type of compute workers, memory, time of execution) needed to reproduce the experiments?
    \item[] Answer: \answerYes{} 
    \item[] Justification: We report the implementation details in Section~\ref{sec:experiments}.
    
\item {\bf Code of ethics}
    \item[] Question: Does the research conducted in the paper conform, in every respect, with the NeurIPS Code of Ethics \url{https://neurips.cc/public/EthicsGuidelines}?
    \item[] Answer: \answerYes{} 
    \item[] Justification: We have reviewed the NeurIPS Code of Ethics and believe the research conforms to it. The work uses public code preference benchmarks and does not collect personally identifiable information or deploy systems for consequential decision-making.

\item {\bf Broader impacts}
    \item[] Question: Does the paper discuss both potential positive societal impacts and negative societal impacts of the work performed?
    \item[] Answer: \answerYes{} 
    \item[] Justification: In Appendix~\ref{app:broader_impact}, we discuss the broader impacts.
    
\item {\bf Safeguards}
    \item[] Question: Does the paper describe safeguards that have been put in place for responsible release of data or models that have a high risk for misuse (e.g., pre-trained language models, image generators, or scraped datasets)?
    \item[] Answer: \answerNA{} 
    \item[] Justification: The paper does not release a high-risk generative model, scraped dataset, or system intended for unrestricted deployment. Our released artifacts, if any, consist of evaluation code, prompts, and derived annotations for research use.

\item {\bf Licenses for existing assets}
    \item[] Question: Are the creators or original owners of assets (e.g., code, data, models), used in the paper, properly credited and are the license and terms of use explicitly mentioned and properly respected?
    \item[] Answer: \answerNo{} 
    \item[] Justification: We cite all existing assets used in the paper, but were unable to identify explicit license information for some benchmark components. We state this limitation and use the assets according to their publicly documented terms.

\item {\bf New assets}
    \item[] Question: Are new assets introduced in the paper well documented and is the documentation provided alongside the assets?
    \item[] Answer: \answerNA{} 
    \item[] Justification: The paper does not introduce or release new datasets, models, or other standalone research assets beyond the method and experimental code.

\item {\bf Crowdsourcing and research with human subjects}
    \item[] Question: For crowdsourcing experiments and research with human subjects, does the paper include the full text of instructions given to participants and screenshots, if applicable, as well as details about compensation (if any)? 
    \item[] Answer: \answerNA{} 
    \item[] Justification: We do not conduct new crowdsourcing experiments or collect new human-subject data. The human preference labels used in our experiments come from existing public benchmarks, which are cited in the paper.

\item {\bf Institutional review board (IRB) approvals or equivalent for research with human subjects}
    \item[] Question: Does the paper describe potential risks incurred by study participants, whether such risks were disclosed to the subjects, and whether Institutional Review Board (IRB) approvals (or an equivalent approval/review based on the requirements of your country or institution) were obtained?
    \item[] Answer: \answerNA{} 
    \item[] Justification: The paper does not involve new crowdsourcing or human-subject experiments conducted by the authors. We only use existing benchmark annotations.

\item {\bf Declaration of LLM usage}
    \item[] Question: Does the paper describe the usage of LLMs if it is an important, original, or non-standard component of the core methods in this research? Note that if the LLM is used only for writing, editing, or formatting purposes and does \emph{not} impact the core methodology, scientific rigor, or originality of the research, declaration is not required.
    \item[] Answer: \answerYes{} 
    \item[] Justification: LLMs are a core component of our method, used for task-specific criterion generation, pairwise criterion judging, optional tie-driven refinement, and final preference prediction.

\end{enumerate}

%% file: main.bib
@article{bigcodearena2025,
  title={BigCodeArena: Unveiling More Reliable Human Preferences in Code Generation via Execution},
  author={Zhuo, Terry Yue and Jin, Xiaolong and Liu, Hange and Jiang, Juyong and Liu, Tianyang and Gong, Chen and Bishnoi, Bhupesh and Mishra, Vaisakhi and Suppa, Marek and Ziems, Noah and others},
  journal={arXiv preprint arXiv:2510.08697},
  year={2025}
}

@article{webdevjudge2025,
  title={WebDevJudge: Evaluating (M) LLMs as critiques for web development quality},
  author={Li, Chunyang and Zheng, Yilun and Huang, Xinting and Fang, Tianqing and Xu, Jiahao and Chen, Lihui and Song, Yangqiu and Hu, Han},
  journal={arXiv preprint arXiv:2510.18560},
  year={2025}
}

@misc{qwen25,
      title={Qwen2.5-VL Technical Report}, 
      author={Shuai Bai and Keqin Chen and Xuejing Liu and Jialin Wang and Wenbin Ge and Sibo Song and Kai Dang and Peng Wang and Shijie Wang and Jun Tang and Humen Zhong and Yuanzhi Zhu and Mingkun Yang and Zhaohai Li and Jianqiang Wan and Pengfei Wang and Wei Ding and Zheren Fu and Yiheng Xu and Jiabo Ye and Xi Zhang and Tianbao Xie and Zesen Cheng and Hang Zhang and Zhibo Yang and Haiyang Xu and Junyang Lin},
      year={2025},
      eprint={2502.13923},
      archivePrefix={arXiv},
      primaryClass={cs.CV},
}

@article{rrd2026,
  title={Rethinking Rubric Generation for Improving LLM Judge and Reward Modeling for Open-ended Tasks},
  author={Shen, William F and Qiu, Xinchi and Whitehouse, Chenxi and Alazraki, Lisa and Goel, Shashwat and Barbieri, Francesco and Willi, Timon and Mathur, Akhil and Leontiadis, Ilias},
  journal={arXiv preprint arXiv:2602.05125},
  year={2026}
}

@inproceedings{rubricisall2025,
  title={Rubric is all you need: Improving llm-based code evaluation with question-specific rubrics},
  author={Pathak, Aditya and Gandhi, Rachit and Uttam, Vaibhav and Ramamoorthy, Arnav and Ghosh, Pratyush and Jindal, Aaryan Raj and Verma, Shreyash and Mittal, Aditya and Ased, Aashna and Khatri, Chirag and others},
  booktitle={Proceedings of the 2025 ACM Conference on International Computing Education Research V. 1},
  pages={181--195},
  year={2025}
}

@article{chasingtail2025,
  title={Chasing the Tail: Effective Rubric-based Reward Modeling for Large Language Model Post-Training},
  author={Zhang, Junkai and Wang, Zihao and Gui, Lin and Sathyendra, Swarnashree Mysore and Jeong, Jaehwan and Veitch, Victor and Wang, Wei and He, Yunzhong and Liu, Bing and Jin, Lifeng},
  journal={arXiv preprint arXiv:2509.21500},
  year={2025}
}

@article{sonnet2024,
  title={Claude 3.5 sonnet model card addendum},
  author={Anthropic, AI},
  journal={Claude-3.5 Model Card},
  volume={3},
  number={6},
  year={2024}
}

@article{kimi2025,
  title={Kimi-vl technical report},
  author={Team, Kimi and Du, Angang and Yin, Bohong and Xing, Bowei and Qu, Bowen and Wang, Bowen and Chen, Cheng and Zhang, Chenlin and Du, Chenzhuang and Wei, Chu and others},
  journal={arXiv preprint arXiv:2504.07491},
  year={2025}
}

@techreport{gemma42026,
  author      = {{Google DeepMind}},
  title       = {Gemma 4: Our most capable open models to date},
  institution = {Google DeepMind},
  year        = {2026},
  url         = {https://deepmind.google/models/gemma/gemma-4/},
  note        = {Technical Report}
}

@inproceedings{vllm2023,
  title={Efficient Memory Management for Large Language Model Serving with PagedAttention},
  author={Woosuk Kwon and Zhuohan Li and Siyuan Zhuang and Ying Sheng and Lianmin Zheng and Cody Hao Yu and Joseph E. Gonzalez and Hao Zhang and Ion Stoica},
  booktitle={Proceedings of the ACM SIGOPS 29th Symposium on Operating Systems Principles},
  year={2023}
}

@article{codejudgebench2025,
  title={Codejudgebench: Benchmarking llm-as-a-judge for coding tasks},
  author={Jiang, Hongchao and Chen, Yiming and Cao, Yushi and Lee, Hung-yi and Tan, Robby T},
  journal={arXiv preprint arXiv:2507.10535},
  year={2025}
}

@article{codeultrafeedback2025,
  title={Codeultrafeedback: An llm-as-a-judge dataset for aligning large language models to coding preferences},
  author={Weyssow, Martin and Kamanda, Aton and Zhou, Xin and Sahraoui, Houari},
  journal={ACM Transactions on Software Engineering and Methodology},
  volume={35},
  number={3},
  pages={1--36},
  year={2026},
  publisher={ACM New York, NY}
}

@inproceedings{prometheus2024,
  title={Prometheus: Inducing fine-grained evaluation capability in language models},
  author={Kim, Seungone and Shin, Jamin and Cho, Yejin and Jang, Joel and Longpre, Shayne and Lee, Hwaran and Yun, Sangdoo and Shin, Seongjin and Kim, Sungdong and Thorne, James and others},
  booktitle={The Twelfth International Conference on Learning Representations},
  year={2024},
}

@article{cai2022,
  title={Constitutional ai: Harmlessness from ai feedback},
  author={Bai, Yuntao and Kadavath, Saurav and Kundu, Sandipan and Askell, Amanda and Kernion, Jackson and Jones, Andy and Chen, Anna and Goldie, Anna and Mirhoseini, Azalia and McKinnon, Cameron and others},
  journal={arXiv preprint arXiv:2212.08073},
  year={2022}
}

@inproceedings{
judgelm2025,
title={Judge{LM}: Fine-tuned Large Language Models are Scalable Judges},
author={Lianghui Zhu and Xinggang Wang and Xinlong Wang},
booktitle={The Thirteenth International Conference on Learning Representations},
year={2025},
}

@article{humaneval2021,
  title={Evaluating large language models trained on code},
  author={Chen, Mark and Tworek, Jerry and Jun, Heewoo and Yuan, Qiming and Pinto, Henrique Ponde De Oliveira and Kaplan, Jared and Edwards, Harri and Burda, Yuri and Joseph, Nicholas and Brockman, Greg and others},
  journal={arXiv preprint arXiv:2107.03374},
  year={2021}
}

@inproceedings{
swebench2024,
title={{SWE}-bench: Can Language Models Resolve Real-world Github Issues?},
author={Carlos E Jimenez and John Yang and Alexander Wettig and Shunyu Yao and Kexin Pei and Ofir Press and Karthik R Narasimhan},
booktitle={The Twelfth International Conference on Learning Representations},
year={2024},
}

@article{llmasjudgesurvey2024,
  title={A survey on llm-as-a-judge},
  author={Gu, Jiawei and Jiang, Xuhui and Shi, Zhichao and Tan, Hexiang and Zhai, Xuehao and Xu, Chengjin and Li, Wei and Shen, Yinghan and Ma, Shengjie and Liu, Honghao and others},
  journal={The Innovation},
  year={2024},
  publisher={Elsevier}
}

@article{llmasjudgesurvey22024,
  title={Llms-as-judges: a comprehensive survey on llm-based evaluation methods},
  author={Li, Haitao and Dong, Qian and Chen, Junjie and Su, Huixue and Zhou, Yujia and Ai, Qingyao and Ye, Ziyi and Liu, Yiqun},
  journal={arXiv preprint arXiv:2412.05579},
  year={2024}
}

@article{llmservey2024l,
  title={Large language models: A survey},
  author={Minaee, Shervin and Mikolov, Tomas and Nikzad, Narjes and Chenaghlu, Meysam and Socher, Richard and Amatriain, Xavier and Gao, Jianfeng},
  journal={arXiv preprint arXiv:2402.06196},
  year={2024}
}

@inproceedings{userstudy_codegen_2024,
  title={User centric evaluation of code generation tools},
  author={Miah, Tanha and Zhu, Hong},
  booktitle={2024 IEEE International Conference on Artificial Intelligence Testing (AITest)},
  pages={109--119},
  year={2024},
  organization={IEEE}
}

@InProceedings{copilotarena2025,
  title = 	 {Copilot Arena: A Platform for Code {LLM} Evaluation in the Wild},
  author =       {Chi, Wayne and Chen, Valerie and Angelopoulos, Anastasios Nikolas and Chiang, Wei-Lin and Mittal, Aditya and Jain, Naman and Zhang, Tianjun and Stoica, Ion and Donahue, Chris and Talwalkar, Ameet},
  booktitle = 	 {Proceedings of the 42nd International Conference on Machine Learning},
  pages = 	 {10354--10382},
  year = 	 {2025},
  editor = 	 {Singh, Aarti and Fazel, Maryam and Hsu, Daniel and Lacoste-Julien, Simon and Berkenkamp, Felix and Maharaj, Tegan and Wagstaff, Kiri and Zhu, Jerry},
  volume = 	 {267},
  series = 	 {Proceedings of Machine Learning Research},
  month = 	 {13--19 Jul},
  publisher =    {PMLR}
}

@article{mctsjudge2025,
  title={Mcts-judge: Test-time scaling in llm-as-a-judge for code correctness evaluation},
  author={Wang, Yutong and Ji, Pengliang and Yang, Chaoqun and Li, Kaixin and Hu, Ming and Li, Jiaoyang and Sartoretti, Guillaume},
  journal={arXiv preprint arXiv:2502.12468},
  year={2025}
}

@article{openrs2026,
  title={Open Rubric System: Scaling Reinforcement Learning with Pairwise Adaptive Rubric},
  author={Jia, Ruipeng and Yang, Yunyi and Wu, Yuxin and Gai, Yongbo and Tao, Siyuan and Zhou, Mengyu and Lin, Jianhe and Jiang, Xiaoxi and Jiang, Guanjun},
  journal={arXiv preprint arXiv:2602.14069},
  year={2026}
}

@article{cdrrm2026,
  title={CDRRM: Contrast-Driven Rubric Generation for Reliable and Interpretable Reward Modeling},
  author={Liu, Dengcan and Yang, Fengkai and Wang, Xiaohan and Yan, Shurui and Chai, Jiajun and Li, Jiahao and Ban, Yikun and Mao, Zhendong and Lin, Wei and Yin, Guojun},
  journal={arXiv preprint arXiv:2603.08035},
  year={2026}
}

@article{trace2026,
  title={Comparing Developer and LLM Biases in Code Evaluation},
  author={Mittal, Aditya and Shar, Ryan and Wu, Zichu and Agarwal, Shyam and Wu, Tongshuang and Donahue, Chris and Talwalkar, Ameet and Chi, Wayne and Chen, Valerie},
  journal={arXiv preprint arXiv:2603.24586},
  year={2026}
}

@inproceedings{codejudge2024,
  title={Codejudge: Evaluating code generation with large language models},
  author={Tong, Weixi and Zhang, Tianyi},
  booktitle={Proceedings of the 2024 Conference on Empirical Methods in Natural Language Processing},
  pages={20032--20051},
  year={2024}
}

@inproceedings{codejudgeeval2024,
  title={CodeJudge-eval: Can large language models be good judges in code understanding?},
  author={Zhao, Yuwei and Luo, Ziyang and Tian, Yuchen and Lin, Hongzhan and Yan, Weixiang and Li, Annan and Ma, Jing},
  booktitle={Proceedings of the 31st International Conference on Computational Linguistics},
  pages={73--95},
  year={2025}
}

@inproceedings{codevisionary2025,
  title={An Agent-based Evaluation Framework for Complex Code Generation},
  author={Wang, Xinchen and Hu, Ruida and Gao, Pengfei and Peng, Chao and Gao, Cuiyun},
  booktitle={2025 40th IEEE/ACM International Conference on Automated Software Engineering (ASE)},
  pages={2427--2439},
  year={2025},
  organization={IEEE}
}

@article{codefavor2024,
  title={Learning code preference via synthetic evolution},
  author={Liu, Jiawei and Nguyen, Thanh and Shang, Mingyue and Ding, Hantian and Li, Xiaopeng and Yu, Yu and Kumar, Varun and Wang, Zijian},
  journal={arXiv preprint arXiv:2410.03837},
  year={2024}
}

@article{vibechecker2025,
  title={Vibe Checker: Aligning Code Evaluation with Human Preference},
  author={Zhong, Ming and Zhou, Xiang and Chang, Ting-Yun and Wang, Qingze and Xu, Nan and Si, Xiance and Garrette, Dan and Upadhyay, Shyam and Liu, Jeremiah and Han, Jiawei and others},
  journal={arXiv preprint arXiv:2510.07315},
  year={2025}
}

@inproceedings{geval2023,
  title={G-eval: NLG evaluation using gpt-4 with better human alignment},
  author={Liu, Yang and Iter, Dan and Xu, Yichong and Wang, Shuohang and Xu, Ruochen and Zhu, Chenguang},
  booktitle={Proceedings of the 2023 conference on empirical methods in natural language processing},
  pages={2511--2522},
  year={2023}
}

@inproceedings{prometheus2_2024,
  title={Prometheus 2: An open source language model specialized in evaluating other language models},
  author={Kim, Seungone and Suk, Juyoung and Longpre, Shayne and Lin, Bill Yuchen and Shin, Jamin and Welleck, Sean and Neubig, Graham and Lee, Moontae and Lee, Kyungjae and Seo, Minjoon},
  booktitle={Proceedings of the 2024 Conference on Empirical Methods in Natural Language Processing},
  pages={4334--4353},
  year={2024}
}

@inproceedings{hdeval2024,
  title={Hd-eval: Aligning large language model evaluators through hierarchical criteria decomposition},
  author={Liu, Yuxuan and Yang, Tianchi and Huang, Shaohan and Zhang, Zihan and Huang, Haizhen and Wei, Furu and Deng, Weiwei and Sun, Feng and Zhang, Qi},
  booktitle={Proceedings of the 62nd Annual Meeting of the Association for Computational Linguistics (Volume 1: Long Papers)},
  pages={7641--7660},
  year={2024}
}

@inproceedings{tickeval2024,
  title={TICKing All the Boxes: Generated Checklists Improve LLM Evaluation and Generation},
  author={Cook, Jonathan and Rockt{\"a}schel, Tim and Foerster, Jakob Nicolaus and Aumiller, Dennis and Wang, Alex},
  booktitle={Language Gamification-NeurIPS 2024 Workshop},
  year={2024}
}

@article{autorubric2026,
  title={Autorubric: Unifying Rubric-based LLM Evaluation},
  author={Rao, Delip and Callison-Burch, Chris},
  journal={arXiv preprint arXiv:2603.00077},
  year={2026}
}

@article{rulers2026,
  title={RULERS: Locked Rubrics and Evidence-Anchored Scoring for Robust LLM Evaluation},
  author={Hong, Yihan and Yao, Huaiyuan and Shen, Bolin and Xu, Wanpeng and Wei, Hua and Dong, Yushun},
  journal={arXiv preprint arXiv:2601.08654},
  year={2026}
}

@inproceedings{iruler2026,
  title={iRULER: Intelligible Rubric-Based User-Defined LLM Evaluation for Revision},
  author={Bai, Jingwen and Cheong, Wei Soon and Muller, Philippe and Lim, Brian Y},
  booktitle={Proceedings of the 2026 CHI Conference on Human Factors in Computing Systems},
  pages={1--28},
  year={2026}
}

@inproceedings{learningtojudge2026,
  title={Learning to judge: LLMs designing and applying evaluation rubrics},
  author={Siro, Clemencia and Aliannejadi, Pourya and Aliannejadi, Mohammad},
  booktitle={Findings of the Association for Computational Linguistics: EACL 2026},
  pages={6371--6389},
  year={2026}
}

@inproceedings{rar2025,
  title={Rubrics as Rewards: Reinforcement Learning Beyond Verifiable Domains},
  author={Gunjal, Anisha and Wang, Anthony and Lau, Elaine and Nath, Vaskar and He, Yunzhong and Liu, Bing and Hendryx, Sean M},
  booktitle={NeurIPS 2025 Workshop on Efficient Reasoning},
  year={2025}
}

@article{onlinerubrics2025,
  title={Online rubrics elicitation from pairwise comparisons},
  author={Rezaei, MohammadHossein and Vacareanu, Robert and Wang, Zihao and Wang, Clinton and Liu, Bing and He, Yunzhong and Aky{\"u}rek, Afra Feyza},
  journal={arXiv preprint arXiv:2510.07284},
  year={2025}
}

@article{openrubrics2025,
  title={Openrubrics: Towards scalable synthetic rubric generation for reward modeling and llm alignment},
  author={Liu, Tianci and Xu, Ran and Yu, Tony and Hong, Ilgee and Yang, Carl and Zhao, Tuo and Wang, Haoyu},
  journal={arXiv preprint arXiv:2510.07743},
  year={2025}
}

@inproceedings{selfpreferencebias2024,
  title={Self-Preference Bias in LLM-as-a-Judge},
  author={Wataoka, Koki and Takahashi, Tsubasa and Ri, Ryokan},
  booktitle={Neurips Safe Generative AI Workshop 2024},
  year={2024}
}

@inproceedings{dontjudgecode2025,
  title={Don’t judge code by its cover: Exploring biases in llm judges for code evaluation},
  author={Moon, Jiwon and Hwang, Yerin and Lee, Dongryeol and Kang, Taegwan and Kim, Yongil and Jung, Kyomin},
  booktitle={Findings of the Association for Computational Linguistics: EACL 2026},
  pages={1364--1389},
  year={2026}
}

@inproceedings{trustorescalate2024,
  title={Trust or Escalate: LLM Judges with Provable Guarantees for Human Agreement},
  author={Jung, Jaehun and Brahman, Faeze and Choi, Yejin},
  booktitle={The Thirteenth International Conference on Learning Representations},
  year={2025},
}

@inproceedings{cloud2024,
  title={Critique-out-Loud Reward Models},
  author={Ankner, Zachary and Paul, Mansheej and Cui, Brandon and Chang, Jonathan Daniel and Ammanabrolu, Prithviraj},
  booktitle={Pluralistic Alignment Workshop at NeurIPS 2024},
  year={2024},
}

@inproceedings{genrm2024,
  title={Generative Verifiers: Reward Modeling as Next-Token Prediction},
  author={Zhang, Lunjun and Hosseini, Arian and Bansal, Hritik and Kazemi, Mehran and Kumar, Aviral and Agarwal, Rishabh},
  booktitle={The Thirteenth International Conference on Learning Representations},
  year={2025},
}

@inproceedings{salmon2023,
  title={SALMON: Self-Alignment with Instructable Reward Models},
  author={Sun, Zhiqing and Shen, Yikang and Zhang, Hongxin and Zhou, Qinhong and Chen, Zhenfang and Cox, David Daniel and Yang, Yiming and Gan, Chuang},
  booktitle={The Twelfth International Conference on Learning Representations},
  year={2024},
}

@inproceedings{icai2025,
  title={Inverse Constitutional AI: Compressing Preferences into Principles},
  author={Findeis, Arduin and Kaufmann, Timo and H{\"u}llermeier, Eyke and Albanie, Samuel and Mullins, Robert D},
  booktitle={The Thirteenth International Conference on Learning Representations},
  year={2025},
}

@article{spct2025,
  title={Inference-time scaling for generalist reward modeling},
  author={Liu, Zijun and Wang, Peiyi and Xu, Runxin and Ma, Shirong and Ruan, Chong and Li, Peng and Liu, Yang and Wu, Yu},
  journal={arXiv preprint arXiv:2504.02495},
  year={2025}
}
